\documentclass[a4paper,10pt]{article}

\usepackage{graphicx}
\usepackage{cite}
\usepackage{a4wide}
\usepackage{amsthm}
\usepackage{amsmath}
\usepackage{amssymb}
\usepackage{caption}

\usepackage{float}
\restylefloat{table}
\usepackage[usenames]{color}

\def \bea { \begin{eqnarray}}
\def \eea {\end{eqnarray}}
\def \be {\begin{equation}}
\def \ee {\end{equation}}

\newcommand{\dd}{{\rm{d}}}
\newcommand{\dD}{{\rm{D}}}
\newcommand{\pul}{\textstyle{\frac{1}{2}}}

\def \bk {\boldsymbol{k}}
\def \bl {\boldsymbol{l}}
\def \bll {\boldsymbol{\ell}}
\def \bm {\boldsymbol{m}}
\def \bn {\boldsymbol{n}}

\def \BE {\begin{equation}}
\def \EE {\end{equation}}
\def \BEA { \begin{eqnarray}}
\def \EEA {\end{eqnarray}}
\def \bea { \begin{eqnarray}}
\def \eea {\end{eqnarray}}
\def \be {\begin{equation}}
\def \ee {\end{equation}}

\title{The Newman--Penrose formalism in quadratic gravity}
\author{ R.~\v{S}varc$^\star$, A.~Pravdov\' a$^\diamond$, D.~Mi{\v s}kovsk\'y$^\star$ \\
\vspace{0.05cm} \\
{\small $^\star$ Institute of Theoretical Physics, Charles University, Prague,} \\
{\small Faculty of Mathematics and Physics, V~Hole\v{s}ovi\v{c}k\'ach~2, 180~00 Praha 8, Czech Republic.} \\
{\small $^\diamond$ Institute of Mathematics of the Czech Academy of Sciences}, \\
{\small \v Zitn\' a 25, 115 67 Prague 1, Czech Republic.}\\[3mm]
{\small E-mail: \texttt{robert.svarc@mff.cuni.cz, pravdova@math.cas.cz, davidmiskovsky@icloud.com }}}
\date{September 29, 2022}

\begin{document}

\maketitle

\begin{abstract}
	The quadratic gravity constraints are reformulated in terms of the Newman--Penrose-like quantities. In such a frame language, the field equations represent a linear algebraic system for the Ricci tensor components. In principle, a procedure for the combination of the Ricci components with standard geometric identities can be applied in a similar way as in the case of general relativity. These results could serve in various subsequent analyses and physical interpretations of admitted solutions to quadratic gravity. Here, we demonstrate the utility of such an approach by explicitly proving general propositions restricting the spacetime geometry under assumptions on a specific algebraic type of curvature tensors.
\end{abstract}

\tableofcontents

\section{Introduction}\label{Sec:Intro}

In 1915, Albert Einstein finished his theory of general relativity (GR) \cite{Einstein15a}, which provided a geometric description of gravity in terms of curved four-dimensional spacetime. Einstein's theory brought dozens of surprising implications during more than a century of its analyses and astrophysical applications.  However, simultaneously with its formulation, there appeared concerns about the possibility of solving its highly complicated nonlinear field equations. These doubts were allayed almost immediately by Karl Schwarzschild and his famous spherically symmetric solution \cite{Schw16}. Unfortunately, the Schwarzschild spacetime also uncovered difficulties related to physical interpretation and insecurity of employing a particular coordinate choice. In the following decades, the construction of coordinate-independent quantities, revealing the true nature of a given gravitational field, become crucial. The conceptually important step within the coordinate-independent analysis was to express studied quantities in terms of their frame components. The privileged role within the frame approaches to Einstein's theory plays the Newman--Penrose (NP) formalism \cite{NewmanPenrose:1962} employing a null vector basis, see subsection~\ref{SubSec:NP} for its summary. The spacetime description in terms of the frame projections allows one to invariantly define the ansatz geometry, e.g., admitting specific null congruences or special algebraic structure of related tensors, and then try to find and discuss its explicit form restricted by the field equations and geometric identities. Importantly, this formalism naturally reflects distinct parts of a gravitational field and its peeling properties. The generalization of NP formalism becomes important also in studies of higher-dimensional GR \cite{Coleyetal04,Durkeeetal10,OrtPraPra13rev,OrtPraPra07}.

Even though GR has beyond doubt proved its ability to describe various strong-field gravitational situations and processes, there remain important theoretical issues unclarified.  One can think, e.g., about its combination with quantum field theory or the nature of singularities that necessarily occurs in its solutions. Attempts to solve these open problems typically consider additional fields or various modifications of GR, see e.g. reviews \cite{Sotiriou:2010,DeFelice:2010,Capozziello:2011,Clifton:2012}.  Alternatively and more pragmatically, one can study modified gravities to analyze the unique position of GR in the space of general metric theories of gravity. From this perspective, the simplest extension of GR corresponds to the quadratic gravity (QG) \cite{Stelle:1977, Stelle:1978, Smilga:2014, Salvio:2018} including all possible curvature squares into the Einstein-Hilbert action, see also subsection~\ref{SubSec:QG}. Such a class of theories may directly solve some of the open problems (however, simultaneously it introduces new ones), or it may be understood as a higher-order correction to GR induced by some unknown final theory.

To better understand QG and its implications on a geometric level, the exact solution analysis becomes important, see, e.g., QG counterparts to the classic Schwarzschild black hole \cite{LuPerkinsPopeStelle:2015,LuPerkinsPopeStelle:2015b,PodolskySvarcPravdaPravdova:2018,SvarcPodolskyPravdaPravdova:2018} or algebraically special geometries \cite{MalekPravda:2011}. It is extremely interesting to compare solutions to QG with those to GR. However, to do so one has to invariantly define the same geometric ansatz, and therefore, the extension of the Newman--Penrose formalism for the case of quadratic gravity seems to be very natural starting point. This is thus the main aim of our contribution. 

The paper is organized as follows. In the introductory section~\ref{Sec:Intro} we summarise concepts of QG and NP formalism. In section~\ref{Sec:QGconstraints}, the NP  form of the QG field equations is derived which represents our main result. Two simple examples of its use are subsequently discussed, see section~\ref{Sec:Applications}. Finally, in appendix~\ref{App:NP} the geometric NP identities are summarized, in appendices~\ref{App:Notation} and~\ref{App:ComparisonHD}, we compare various conventions related to the NP formalism, and in the last appendix~\ref{App:FullFEs}, the QG field equations are listed in the fully explicit form.
%%%%%%%%%%%%%%%%%%%%%%%%%%%%%%%%%%%%%%%%%%%%%%%%%%%%%%%%%%%%%%%%%%%%%%%%%%%%%%%%%

\subsection{Quadratic gravity}
\label{SubSec:QG}

The vacuum quadratic gravity can be introduced via its action as
\begin{equation}
S = \int \left[\frac{1}{\mathsf{k}}(R-2\Lambda)-\mathfrak{a}\,C_{abcd}C^{abcd}+\mathfrak{b}\,R^2\right]\sqrt{-g}\dd^{4}x \,, \label{QG_action}
\end{equation}
where $R$ is the Ricci scalar, ${C_{abcd}}$ is the Weyl tensor, $\mathsf{k}$, $\mathfrak{a}$ and $\mathfrak{b}$ are coupling constants of the theory, and $\Lambda$ stands for the cosmological term, see, e.g., \cite{Stelle:1977, Stelle:1978, Smilga:2014, Salvio:2018}. Due to the Gauss--Bonnet theorem, this represents the most general class of four-dimensional quadratic theories. Subsequently, the least action principle ${\delta S=0}$ leads to the 4th order field equations in the form
\begin{equation}
\frac{1}{\mathsf{k}}\left(R_{ab}-\frac{1}{2}R g_{ab}+\Lambda g_{ab}\right)-4\mathfrak{a}\,B_{ab}+2\mathfrak{b}\left(R_{ab}-\frac{1}{4}R g_{ab}+g_{ab}\Box-\nabla_{a}\nabla_{b}\right)R=0 \,, \label{QG_FEqs}
\end{equation}
where $B_{ab}$ is the Bach tensor defined as
\begin{equation}
B_{ab} = \left(\nabla^{c}\nabla^{d}+\frac{1}{2}R^{cd}\right)C_{acbd} \,, \label{Bach_definition}
\end{equation}
which is symmetric, trace-less, covariantly constant, and conformaly re-scaled, i.e.,
\begin{equation}
B_{ab} = B_{ba} \,, \quad B_{ab}g^{ab} = 0 \,, \quad B_{ab;c}\,g^{bc} = 0 \,, \quad \tilde{g}_{ab} = \Omega^2 g_{ab} \ \Rightarrow \ \tilde{B}_{ab} = \Omega^{-2} B_{ab} \,.
\end{equation}
The field equations trace yields immediately the condition for the scalar curvature, namely
\begin{equation}
R = 6 \mathfrak{b} \mathsf{k} \Box R + 4 \Lambda \,.
\end{equation}
To employ the Newman--Penrose-like approach to the discussion of admissible gravitational fields in quadratic gravity we separate the Ricci tensor contribution in (\ref{QG_FEqs}). Substituting the Bach tensor (\ref{Bach_definition}) and grouping all terms with the Ricci tensor we thus get
\begin{equation}
\left(\frac{1}{\mathsf{k}} + 2 \mathfrak{b} R \right)R_{ab} - 2 \mathfrak{a} R^{cd}C_{acbd} + Z_{ab} = 0 \,, \label{QG_FEqs_separated_form}
\end{equation}
where $Z_{ab}$ is a shorthand for
\begin{equation}
Z_{ab} = -\frac{1}{\mathsf{k}}\left(\frac{1}{2}R g_{ab}-\Lambda g_{ab}\right)-4\mathfrak{a}\,\nabla^{c}\nabla^{d}C_{acbd}-2\mathfrak{b}\left(\frac{1}{4}R g_{ab}-g_{ab}\Box +\nabla_{a}\nabla_{b}\right)R \,. \label{Z_tensor}
\end{equation}

\subsection{The Newman--Penrose quantities}
\label{SubSec:NP}

To set up the notation and fix the conventions, we summarize essential definitions of the Newman--Penrose formalism.\footnote{Here we follow the notation of classic book \cite{Stephanietal:2003} while its relation to other common sources \cite{Chandrasekhar:1993, PenroseRindler:1984} is summarized in the appendix \ref{App:Notation}.} Subsequently, the geometric constraints including the commutation relations, the Ricci and Bianchi identities are listed in the appendix~\ref{App:NP}.  Let us introduce the null orthonormal frame ${\{\bk\,, \bl\,, \bm\,, \bar{\bm}\}}$, where $\bk$, $\bl$ are real null vectors and $\bm$, $\bar{\bm}$ are complex null vectors, respectively. They are normalized as
\begin{align}
\bk\cdot\bl = -1 \,, \qquad \bm\cdot\bar{\bm} = 1\,, \label{null_condition_NP}
\end{align}
with other combinations being vanishing. The metric thus becomes
\begin{equation}
g_{ab}= -2k_{(a}l_{b)}+2m_{(a}\bar{m}_{b)}\,.
\end{equation}
Freedom in a choice of the frame is given by the Lorentz transformations, namely
\begin{itemize}
\item boost in the plane of null vectors $\bk$ and $\bl$ with a positive parameter $A$:
\begin{equation}
k^{a} \mapsto A\,k^{a}\,, \qquad l^{a} \mapsto A^{-1}\,l^{a}\,, \qquad m^{a} \mapsto m^{a}\,,
\end{equation}
\item rotation in the transverse space of vectors $\boldsymbol{m}$ and $\bar{\boldsymbol{m}}$ encoded in a real parameter $\Theta$:
\begin{equation}
k^{a} \mapsto k^{a}\,, \qquad l^{a} \mapsto l^{a}\,, \qquad m^{a} \mapsto e^{\mathrm{i}\Theta}\,m^{a}\,,
\end{equation}
\item null rotation with $\bk$ fixed given by a complex parameter $B$:
\begin{equation}
k^{a} \mapsto k^{a}\,, \qquad m^{a} \mapsto m^{a} + B\,k^{a}\,, \qquad l^{a} \mapsto l^{a} + \bar{B}\,m^{a} + B\,\bar{m}^{a} + |B|^2\,k^{a}\,,
\end{equation}
\item null rotation with $\bl$ fixed given by a complex parameter $E$:
\begin{equation}
l^{a} \mapsto{l}^{a}\,, \qquad m^{a} \mapsto {m}^{a} + E\,{l}^{a}\,, \qquad k^{a} \mapsto k^{a} + \bar{E}\,{m}^{a} + E\,\bar{m}^{a} + |E|^2\,l^{a}\,.
\end{equation}
\end{itemize}
The covariant derivative components in the frame vector directions are denoted by
\begin{align}
\dD = k^{a}\nabla_{a} \,, \quad \Delta = l^{a}\nabla_{a} \,, \quad \delta = m^{a}\nabla_{a} \,, \quad \bar{\delta} = \bar{m}^{a}\nabla_{a} \,. \label{NP_formalism_derivatives}
\end{align}
To characterize the above derivatives acting of the frame vectors we define the spin coefficients as
\begin{align}
\kappa&= -k_{a;b}m^{a}k^{b} \,,& \quad \nu&= l_{a;b}\bar{m}^{a}l^{b} \,,& \quad \epsilon&= \frac{1}{2}\left(m_{a;b}\bar{m}^{a}k^{b}-k_{a;b}l^{a}k^{b}\right) , \nonumber \\
\rho&= -k_{a;b}m^{a}\bar{m}^{b} \,,& \quad \mu&= l_{a;b}\bar{m}^{a}m^{b} \,,& \quad \beta&= \frac{1}{2}\left(m_{a;b}\bar{m}^{a}m^{b}-k_{a;b}l^{a}m^{b}\right) , \nonumber \\
\sigma&= -k_{a;b}m^{a}m^{b} \,,& \quad \lambda&= l_{a;b}\bar{m}^{a}\bar{m}^{b} \,,& \quad \gamma&= \frac{1}{2}\left(l_{a;b}k^{a}l^{b}-\bar{m}_{a;b}m^{a}l^{b}\right) , \nonumber \\
\tau&= -k_{a;b}m^{a}l^{b} \,,& \quad \pi&= l_{a;b}\bar{m}^{a}k^{b} \,,& \quad \alpha&= \frac{1}{2}\left(l_{a;b}k^{a}\bar{m}^{b}-\bar{m}_{a;b}m^{a}\bar{m}^{b}\right) . \label{NP_formalism_spin_coefficients}
\end{align}

The Weyl tensor null tetrad independent components are
\begin{align}
    \Psi_{0}&= C_{abcd} k^{a} m^{b} k^{c} m^{d}\,, \nonumber \\
    \Psi_{1}&= C_{abcd} k^{a} l^{b} k^{c} m^{d}\,, \nonumber \\
    \Psi_{2}&= C_{abcd} k^{a} m^{b} \bar{m}^{c} l^{d}=\frac{1}{2}C_{abcd}k^{a}l^{b}(k^{c}l^{d}-m^{c}\bar{m}^{d})\,, \nonumber \\
    \Psi_{3}&= C_{abcd} l^{a} k^{b} l^{c} \bar{m}^{d}\,, \nonumber \\
    \Psi_{4}&= C_{abcd} l^{a} \bar{m}^{b} l^{c} \bar{m}^{d}\,,
    \label{Weyl_components_NP}
\end{align}
and the projections of the Ricci tensor (or equivalently its traceless part ${S_{ab}=R_{ab}-\frac{1}{4}Rg_{ab}}$) can be introduced as
\begin{align}
    \Phi_{00} & = \frac{1}{2} R_{ab}k^{a}k^{b} \,,& & & &\nonumber \\
    \Phi_{01} & = \frac{1}{2} R_{ab}k^{a}m^{b} \,,& \quad \Phi_{10} & = \frac{1}{2} R_{ab}k^{a}\bar{m}^{b}\,, & & \nonumber \\
    \Phi_{11} & = \frac{1}{4}R_{ab}\left(k^{a}l^{b} + m^{a}\bar{m}^{b}\right) , & \quad \Phi_{02} & = \frac{1}{2} R_{ab}m^{a}m^{b} \,,& \quad \Phi_{20} & = \frac{1}{2} R_{ab}\bar{m}^{a}\bar{m}^{b} \,, \nonumber \\
		\Phi_{12} & = \frac{1}{2} R_{ab}l^{a}m^{b} \,, & \quad \Phi_{21} & = \frac{1}{2} R_{ab}l^{a}\bar{m}^{b} \,,& & \nonumber \\
    \Phi_{22} & = \frac{1}{2} R_{ab}l^{a}l^{b} \,,& & & & \label{Ricci_components_NP}
\end{align}
with the trace ${R= 2R_{ab}\left(m^{a}\bar{m}^{b}-k^{a}l^{b}\right)}$ which implies
\begin{equation}
R_{ab}k^{a}l^{b}= -\frac{1}{4}R+2\Phi_{11}\,, \qquad R_{ab}m^{a}\bar{m}^{b}=\frac{1}{4}R+2\Phi_{11} \,.
\end{equation}

\newpage

\section{Quadratic gravity constraints}\label{Sec:QGconstraints}

The quadratic gravity field equations (\ref{QG_FEqs_separated_form}), expressed in terms of the null frame ${\{\bk\,, \bl\,, \bm\,, \bar{\bm}\}}$, take the form 
\begin{align}
0 =& -4\mathfrak{a} \left[\Phi_{20} \Psi_{0} + \Phi_{02} \bar{\Psi}_{0} - 2 \Phi_{10} \Psi_{1} - 2 \Phi_{01} \bar{\Psi}_{1} + \Phi_{00}( \Psi_{2} +  \bar{\Psi}_{2})\right] \nonumber \\
		&+ 2\left(\frac{ 1}{\mathsf{k}} + 2\mathfrak{b}R \right)\Phi_{00} + Z_{(0)(0)} \,, \label{QG_00}\\
0 =& -4\mathfrak{a} \left[\Phi_{21} \Psi_{1} + \Phi_{12} \bar{\Psi}_{1} - 2 \Phi_{11} (\Psi_{2} + \bar{\Psi}_{2}) + \Phi_{01} \Psi_{3} + \Phi_{10} \bar{\Psi}_{3}\right] \nonumber \\ 
    &+ \left(\frac{ 1}{\mathsf{k}} + 2\mathfrak{b}R \right)\left(2 \Phi_{11} - \frac{R}{4}\right) + Z_{(0)(1)}\label{QG_01} \,, \\
0 =& -4\mathfrak{a} \left[\Phi_{21} \Psi_{0} - 2 \Phi_{11} \Psi_{1} + \Phi_{02} \bar{\Psi}_{1} + \Phi_{01}( \Psi_{2} - 2  \bar{\Psi}_{2}) + \Phi_{00} \bar{\Psi}_{3}\right] \nonumber \\
    &+ 2\left(\frac{ 1}{\mathsf{k}} + 2\mathfrak{b}R \right)\Phi_{01} + Z_{(0)(2)} \,, \label{QG_02}\\
0 =& -4\mathfrak{a} \bigl[\Phi_{22} (\Psi_{2} + \bar{\Psi}_{2}) - 2 \Phi_{12} \Psi_{3} - 2 \Phi_{21} \bar{\Psi}_{3} + \Phi_{02} \Psi_{4} + \Phi_{20} \bar{\Psi}_{4}\bigr] \nonumber \\ 
    &+ 2\left(\frac{ 1}{\mathsf{k}} + 2\mathfrak{b}R \right)\Phi_{22} + Z_{(1)(1)} \,, \label{QG_11}\\
0 =& -4\mathfrak{a} \left[\Phi_{22} \Psi_{1}+\Phi_{12}( - 2  \Psi_{2} +  \bar{\Psi}_{2}) + \Phi_{02} \Psi_{3} - 2 \Phi_{11} \bar{\Psi}_{3} + \Phi_{10} \bar{\Psi}_{4}\right] \nonumber \\ 
    &+ 2\left(\frac{ 1}{\mathsf{k}} + 2\mathfrak{b}R \right)\Phi_{12} + Z_{(1)(2)} \,,\label{QG_12} \\
0 =& -4\mathfrak{a} \left[\Phi_{22} \Psi_{0} - 2 \Phi_{12} \Psi_{1} + \Phi_{02} (\Psi_{2} +  \bar{\Psi}_{2}) - 2 \Phi_{01} \bar{\Psi}_{3} + \Phi_{00} \bar{\Psi}_{4}\right] \nonumber \\ 
    &+ 2\left(\frac{ 1}{\mathsf{k}} + 2\mathfrak{b}R \right)\Phi_{02} + Z_{(2)(2)} \,, \label{QG_22}\\
0 =& -4\mathfrak{a} \left[\Phi_{21} \Psi_{1} + \Phi_{12} \bar{\Psi}_{1} - 2 \Phi_{11} (\Psi_{2} + \bar{\Psi}_{2}) + \Phi_{01} \Psi_{3} + \Phi_{10} \bar{\Psi}_{3}\right] \nonumber \\ 
    &+ \left(\frac{ 1}{\mathsf{k}} + 2\mathfrak{b}R \right)\left(2\Phi_{11} + \frac{R}{4}\right) + Z_{(2)(3)} \,,
    \label{QG_23}
     %\label{QG_field_equations}
\end{align}
where components of the Weyl and Ricci tensors are defined by (\ref{Weyl_components_NP}) and (\ref{Ricci_components_NP}), respectively. The symbols ${Z_{(c)(d)} = Z_{ab}e^{\ a}_{(c)}e^{\ b}_{(d)}}$ stand for the frame components of ${Z_{ab}}$ given by (\ref{Z_tensor}), e.g., $Z_{(0)(0)} = Z_{ab}k^{a}k^{b}$ and $Z_{(1)(2)} = Z_{ab}l^{a}m^{b}$ etc. In principle, the above system of equations can be understood as algebraic constraints on the Ricci tensor components which have to be further combined with the geometric conditions\footnote{In fact, the same approach is applied in the context of vacuum Einstein's general relativity, where the Ricci tensor components are also directly restricted by the field equations. However, in such a case (${\mathfrak{a}=0=\mathfrak{b}}$) the constraints are very simple with all components (\ref{Ricci_components_NP}) vanishing and ${R=4\Lambda}$.} listed in appendix~\ref{App:NP}.

Finally, to be fully explicit we express all relevant projections of the $Z_{ab}$ tensor, i.e., 
\begin{align}
Z_{(0)(0)} =& -4\mathfrak{a}B^{Z}_{(0)(0)} + 2\mathfrak{b}\left[(\epsilon + \bar{\epsilon}) \text{D} R - \text{D}\text{D} R - \bar{\kappa} \delta R - \kappa \bar{\delta} R \right], \label{Z_proj_00} \\
Z_{(0)(1)} =& -4\mathfrak{a}B^{Z}_{(0)(1)}+\frac{1}{2\mathsf{k}}(R-2\Lambda) + 2\mathfrak{b}\Big[\,\frac{1}{4} R ^2 - (\gamma + \bar{\gamma} - \mu - \bar{\mu}) \text{D} R \nonumber \\ 
&\quad - (\rho  + \bar{\rho} )\Delta R  + \Delta\text{D} R + (\alpha  - \bar{\beta}  + \bar{\tau} )\delta R - \delta\bar{\delta} R \nonumber \\
&\quad + (\bar{\alpha}  - \beta   + \tau) \bar{\delta} R - \bar{\delta}\delta R\Big]\,, \label{Z_proj_01}\\
Z_{(0)(2)} =& -4\mathfrak{a}B^{Z}_{(0)(2)} + 2\mathfrak{b}\left[\bar{\pi} \text{D} R  -\text{D}\delta R  - \kappa \Delta R  + (\epsilon  - \bar{\epsilon} )\delta R \right], \label{Z_proj_02}\\
Z_{(1)(1)} =& -4\mathfrak{a}B^{Z}_{(1)(1)} + 2\mathfrak{b}\left[-(\gamma + \bar{\gamma}) \Delta R  - \Delta\Delta R  +  \nu \delta R  +  \bar{\nu} \bar{\delta} R  \right]\,, \label{Z_proj_03}\\
Z_{(1)(2)} =& -4\mathfrak{a}B^{Z}_{(1)(2)} + 2\mathfrak{b}\left[\bar{\nu} \text{D} R  -  \tau \Delta R  -  \Delta\delta R  + (\gamma   -  \bar{\gamma} )\delta R \right]\,, \label{Z_proj_12}\\
Z_{(2)(2)} =& -4\mathfrak{a}B^{Z}_{(2)(2)} + 2\mathfrak{b}\left[\bar{\lambda} \text{D} R  -  \sigma \Delta R +( -  \bar{\alpha}   + \beta )\delta R  -  \delta\delta R \right]\,, \label{Z_proj_22} \\
Z_{(2)(3)} =& -4\mathfrak{a}B^{Z}_{(2)(3)}-\frac{1}{2\mathsf{k}}(R-2\Lambda)  + 2\mathfrak{b}\Big[\,-\frac{1}{4} R^2 + (\gamma + \bar{\gamma} -  \bar{\mu}) \text{D} R \nonumber \\ 
&\quad - \text{D}\Delta R +( \rho-\epsilon   - \bar{\epsilon}   ) \Delta R  - \Delta\text{D} R +(- \alpha   + \bar{\beta}+ \pi- \bar{\tau} )\delta R \nonumber \\
&\quad     + (\bar{\pi} - \tau )\bar{\delta} R  + \bar{\delta}\delta R \Big]\,, \label{Z_proj_23}
\end{align}
where ${B^{Z}_{(c)(d)} = B^{Z}_{ab}e^{\ a}_{(c)}e^{\ b}_{(d)}}$ represents the Ricci-independent part of the Bach tensor corresponding to the second covariant derivative of the Weyl tensor, namely
\begin{equation}
    B^{Z}_{ab} = \nabla^{c}\nabla^{d}C_{acbd}\,.
\end{equation}
Explicitly we get 
\begin{align}
    B^{Z}_{(0)(0)}= & \bar{\delta}\bar{\delta}\Psi_{0} - \text{D}\bar{\delta}\Psi_{1} - \bar{\delta}\text{D}\Psi_{1} + \text{D}\text{D}\Psi_{2}
    % \nonumber \\
    %&
    +\lambda \text{D}\Psi_{0} + \bar{\sigma} \Delta\Psi_{0} + ( 2 \pi-7 \alpha -  \bar{\beta} ) \bar{\delta}\Psi_{0} \nonumber \\
    &+(5 \alpha + \bar{\beta} - 3 \pi) \text{D}\Psi_{1} -  \bar{\kappa} \Delta\Psi_{1} -  \bar{\sigma} \delta\Psi_{1} + (3 \epsilon + \bar{\epsilon} + 7 \rho) \bar{\delta}\Psi_{1} \nonumber \\
    &-( \epsilon +  \bar{\epsilon} + 6 \rho) \text{D}\Psi_{2} + \bar{\kappa} \delta\Psi_{2} - 5 \kappa \bar{\delta}\Psi_{2} +4 \kappa \text{D}\Psi_{3} \nonumber \\
    &+\Psi_{0} [\bar{\kappa} \nu+4\alpha( 3 \alpha +  \bar{\beta}) -  (\epsilon+  \bar{\epsilon} + 3  \rho) \lambda  + \pi(\pi- 7 \alpha  -  \bar{\beta} )  +\bar{\sigma}( \mu - 4 \gamma) 
     %\nonumber \\
    %&\hspace{10.0mm}  
    + \text{D}\lambda - 4 \bar{\delta}\alpha + \bar{\delta}\pi] \nonumber \\
    &+2 \Psi_{1} [2 \kappa \lambda+ \bar{\kappa}(\gamma- \mu )+\rho (5\pi-9 \alpha  - 2 \bar{\beta})  + \bar{\sigma}(\beta  + 2 \tau) + \epsilon (2 \pi-4 \alpha - \bar{\beta}) +  \bar{\epsilon}(\pi-\alpha )   \nonumber \\
    &\hspace{14.0mm}  +  \text{D}\alpha - \text{D}\pi +  \bar{\delta}\epsilon + 2 \bar{\delta}\rho] \nonumber \\
    &+3 \Psi_{2} [\kappa(3 \alpha  + \bar{\beta}  - 3  \pi ) -  \bar{\kappa} \tau + \rho (\epsilon + \bar{\epsilon}  + 3 \rho) -  \sigma \bar{\sigma} -  \text{D}\rho -  \bar{\delta}\kappa] \nonumber \\
    &+2 \Psi_{3} [ \kappa(\epsilon -  \bar{\epsilon}  - 5  \rho )+ \bar{\kappa} \sigma + \text{D}\kappa] 
    +2 \Psi_{4} \kappa^2 + c.c.\,, \label{BZ_proj_00}
\end{align}
\begin{align}
B^{Z}_{(0)(1)}= & \bar{\delta}\Delta\Psi_{1} - \text{D}\Delta\Psi_{2} -  \bar{\delta}\delta\Psi_{2} + \text{D}\delta\Psi_{3} - \lambda \Delta\Psi_{0} -  \nu \bar{\delta}\Psi_{0} \nonumber \\
    &+2 \nu \text{D}\Psi_{1} + (2 \pi- \alpha + \bar{\beta} ) \Delta\Psi_{1} + \lambda \delta\Psi_{1} + (2 \mu -  \bar{\mu} -2 \gamma ) \bar{\delta}\Psi_{1} \nonumber \\
    &+(\bar{\mu} -3 \mu ) \text{D}\Psi_{2} + (2 \rho- \epsilon -  \bar{\epsilon} ) \Delta\Psi_{2} + (\alpha -  \bar{\beta} - 2 \pi) \delta\Psi_{2} + (\bar{\pi} + 3 \tau) \bar{\delta}\Psi_{2} \nonumber \\
    &+(2 \beta -  \bar{\pi} - 2 \tau) \text{D}\Psi_{3} -  \kappa \Delta\Psi_{3} + (\epsilon + \bar{\epsilon} - 2 \rho) \delta\Psi_{3} - 2 \sigma \bar{\delta}\Psi_{3} +\sigma \text{D}\Psi_{4} + \kappa \delta\Psi_{4} \nonumber \\
    &+\Psi_{0} [\lambda( 4 \gamma  -   \mu +  \bar{\mu}) + \nu(\alpha  -  \bar{\beta}  - 2  \pi )-  \bar{\delta}\nu] \nonumber \\
    &+2\Psi_{1} [\gamma( \alpha  -  \bar{\beta}  - 2  \pi) -  \lambda(\beta 
    +  \bar{\pi}+2  \tau)+\mu(\bar{\beta} -  \alpha  +   2 \pi) +  \bar{\mu}(\alpha  -  \pi ) +\nu( \epsilon  +  \bar{\epsilon}- 2  \rho ) \nonumber \\
    &\hspace{10.0mm}  +  \text{D}\nu -  \bar{\delta}\gamma +  \bar{\delta}\mu] \nonumber \\
    &+3 \Psi_{2} [  \kappa \nu+\mu( 2  \rho -\epsilon  - \bar{\epsilon} ) - \bar{\mu} \rho +  \pi \bar{\pi}  +  \lambda \sigma + \tau(2 \pi-\alpha  +  \bar{\beta}   ) - \text{D}\mu +  \bar{\delta}\tau] \nonumber \\
    &+2 \Psi_{3} [\kappa (\bar{\mu}- 2  \mu -  \gamma) +\epsilon(\beta -   \tau  -   \bar{\pi}) + \bar{\epsilon} (\beta -  \tau )
    + \rho( \bar{\pi}  - 2 \beta+ 2  \tau) +  \sigma(\alpha  -  \bar{\beta} - 2 \pi) \nonumber \\
    &\hspace{14.0mm}    + \text{D}\beta -  \text{D}\tau -  \bar{\delta}\sigma] \nonumber \\
    &+\Psi_{4} [\kappa(4 \beta  -  \bar{\pi} -   \tau) +  \sigma (\epsilon+ \bar{\epsilon}  - 2 \rho  )+ \text{D}\sigma ] + c.c.\,, \label{BZ_proj_01}
\end{align}
\begin{align}
B^{Z}_{(0)(2)} =& \bar{\delta}\Delta\Psi_{0} - \text{D}\Delta\Psi_{1} -  \bar{\delta}\delta\Psi_{1} + \text{D}\delta\Psi_{2} \nonumber \\
&+\nu \text{D}\Psi_{0} + (\pi-3 \alpha + \bar{\beta} ) \Delta\Psi_{0} + (\mu -  \bar{\mu} -4 \gamma ) \bar{\delta}\Psi_{0} \nonumber \\
&+(2 \gamma - 2 \mu + \bar{\mu}) \text{D}\Psi_{1} + (\epsilon -  \bar{\epsilon} + 3 \rho) \Delta\Psi_{1} + (3 \alpha -  \bar{\beta} -  \pi) \delta\Psi_{1}  + (2 \beta + \bar{\pi} + 4 \tau) \bar{\delta}\Psi_{1} \nonumber \\
&-( \bar{\pi} + 3 \tau) \text{D}\Psi_{2} - 2 \kappa \Delta\Psi_{2} - ( \epsilon - \bar{\epsilon} + 3 \rho) \delta\Psi_{2} - 3 \sigma \bar{\delta}\Psi_{2} +2 \sigma \text{D}\Psi_{3} + 2 \kappa \delta\Psi_{3} \nonumber \\
&+\Psi_{0} [(4\gamma-\mu) (3 \alpha  -  \bar{\beta}-  \pi)   + \bar{\mu}(4 \alpha -  \pi) +\nu( \bar{\epsilon} -  \epsilon - 3 \rho   )
 -  \lambda \bar{\pi}   
 % \nonumber \\
%&\hspace{10.0mm}
+ \text{D}\nu - 4 \bar{\delta}\gamma + \bar{\delta}\mu] \nonumber \\
&+2 \Psi_{1} [ 2 \kappa \nu +(\mu- \gamma) (\epsilon -  \bar{\epsilon} + 3  \rho) -  \bar{\mu}(2 \rho + \epsilon)  + ( \beta+2\tau) (\pi   -3 \alpha  +   \bar{\beta}) 
+  \bar{\pi}(\pi - \alpha) \nonumber \\ 
   &\hspace{14.0mm}
  %\nonumber \\ 
  %   &\hspace{14.0mm}
  +  \text{D}\gamma - \text{D}\mu +  \bar{\delta}\beta + 2 \bar{\delta}\tau] \nonumber \\
    &+3 \Psi_{2} [ \kappa(\bar{\mu} -2 \mu ) +  \bar{\pi} \rho + \sigma(3 \alpha  - \bar{\beta}  - \pi ) + \tau(\epsilon  - \bar{\epsilon}  + 3 \rho ) - \text{D}\tau - \bar{\delta}\sigma] \nonumber \\
    &+2 \Psi_{3} [\kappa(2 \beta  -   \bar{\pi}- 2  \tau) +\sigma(\bar{\epsilon} -  \epsilon   - 3 \rho)  + \text{D}\sigma] + 2 \Psi_{4} \kappa \sigma \nonumber \\
    &+ \delta\delta\bar{\Psi}_{1} - \delta\text{D}\bar{\Psi}_{2} - \text{D}\delta\bar{\Psi}_{2} + \text{D}\text{D}\bar{\Psi}_{3} \nonumber \\
    &- 2 \bar{\lambda} \delta\bar{\Psi}_{0} + 3 \bar{\lambda} \text{D}\bar{\Psi}_{1} +  \sigma \Delta\bar{\Psi}_{1} +(4 \bar{\pi} - 3 \bar{\alpha} - \beta ) \delta\bar{\Psi}_{1} \nonumber \\
    &+( \bar{\alpha} +  \beta - 5 \bar{\pi}) \text{D}\bar{\Psi}_{2} - \kappa \Delta\bar{\Psi}_{2} + ( \epsilon - \bar{\epsilon} + 5 \bar{\rho}) \delta\bar{\Psi}_{2} - \sigma \bar{\delta}\bar{\Psi}_{2} \nonumber \\
    &+( 3 \bar{\epsilon}-\epsilon  - 4 \bar{\rho}) \text{D}\bar{\Psi}_{3} - 3 \bar{\kappa} \delta\bar{\Psi}_{3} +  \kappa \bar{\delta}\bar{\Psi}_{3} + 2 \bar{\kappa} \text{D}\bar{\Psi}_{4} \nonumber \\
    &+\bar{\Psi}_{0} [\bar{\lambda}(5 \bar{\alpha}  +  \beta  - 3  \bar{\pi}) - \bar{\nu} \sigma - \delta\bar{\lambda}] \nonumber \\
    &+2 \bar{\Psi}_{1} [\kappa \bar{\nu}+ \bar{\alpha}(\bar{\alpha} +  \beta ) +\bar{\pi}( 2 \bar{\pi} - 3 \bar{\alpha}  -  \beta)   - \bar{\lambda}( 4  \bar{\rho}+  \epsilon ) + \sigma(\bar{\mu}  -  \bar{\gamma} )  + \text{D}\bar{\lambda} -  \delta\bar{\alpha} + \delta\bar{\pi} ] \nonumber \\
    &+3 \bar{\Psi}_{2} [2 \bar{\kappa} \bar{\lambda} -  \kappa \bar{\mu} +  \bar{\pi}(\epsilon -  \bar{\epsilon})  + \bar{\rho}( 4 \bar{\pi}-  \bar{\alpha}  -  \beta )  + \sigma \bar{\tau} -  \text{D}\bar{\pi} + \delta\bar{\rho}] \nonumber \\
    &+2 \bar{\Psi}_{3} (\kappa  (\bar{\beta}  - \bar{\tau} ) +  \bar{\kappa} (\beta   - 4 \bar{\pi} ) - \sigma \bar{\sigma}
     +  (\bar{\rho} - \bar{\epsilon}) (\epsilon  -  \bar{\epsilon}  + 2 \bar{\rho})  +  \text{D}\bar{\epsilon} - \text{D}\bar{\rho} - \delta\bar{\kappa}) \nonumber \\
    &+\bar{\Psi}_{4} [ \bar{\kappa}(  5 \bar{\epsilon} -\epsilon   - 3  \bar{\rho}) +  \kappa \bar{\sigma} +  \text{D}\bar{\kappa} ]\,, \label{BZ_proj_02}
\end{align}
\begin{align}
B^{Z}_{(1)(2)} = & \Delta\Delta\Psi_{1} -  \Delta\delta\Psi_{2} -  \delta\Delta\Psi_{2} + \delta\delta\Psi_{3} \nonumber \\
    &- 2 \nu \Delta\Psi_{0} + ( 4 \mu-3 \gamma + \bar{\gamma} ) \Delta\Psi_{1} + 3 \nu \delta\Psi_{1} -  \bar{\nu} \bar{\delta}\Psi_{1} \nonumber \\
    &+ \bar{\nu} \text{D}\Psi_{2} + (5 \tau- \bar{\alpha} -  \beta ) \Delta\Psi_{2} + (\gamma -  \bar{\gamma} - 5 \mu) \delta\Psi_{2} + \bar{\lambda} \bar{\delta}\Psi_{2} \nonumber \\
    &- \bar{\lambda} \text{D}\Psi_{3} - 3 \sigma \Delta\Psi_{3} + (\bar{\alpha} + 3 \beta - 4 \tau) \delta\Psi_{3} + 2 \sigma \delta\Psi_{4} \nonumber \\
    &+ \Psi_{0} [\nu ( 5 \gamma  -  \bar{\gamma}  - 3 \mu ) + \lambda \bar{\nu} -  \Delta\nu ] \nonumber \\
    &+ 2 \Psi_{1} [\nu ( \bar{\alpha} - 4  \tau)  + \bar{\nu}(\alpha  -   \pi)
    -  \lambda \bar{\lambda} +(\gamma-\mu) (\gamma -   \bar{\gamma}  - 2  \mu) %\nonumber \\
    %&\hspace{14.0mm} 
    - \Delta\gamma + \Delta\mu + \delta\nu ] \nonumber \\
    &+3\Psi_{2} [ \mu( 4  \tau-  \bar{\alpha}  -  \beta )+  \bar{\lambda} \pi -  \bar{\nu} \rho + 2 \nu \sigma+  \tau ( \bar{\gamma}   -  \gamma )   + \Delta\tau - \delta\mu ] \nonumber \\
    &+ 2 \Psi_{3} [  \kappa \bar{\nu} -   \sigma (\bar{\gamma} + 4 \mu) +\tau ( 2\tau -  \bar{\alpha} - 3 \beta ) + \beta (\bar{\alpha}  + \beta)+ \bar{\lambda} (\rho  -  \epsilon)  -  \Delta\sigma + \delta\beta -  \delta\tau ] \nonumber \\
    &+\Psi_{4} [- \kappa \bar{\lambda} + \sigma (\bar{\alpha} + 5 \beta - 3 \tau) + \delta\sigma ] \nonumber \\
    &- \Delta\text{D}\bar{\Psi}_{3} +  \Delta\delta\bar{\Psi}_{2} +  \bar{\delta}\text{D}\bar{\Psi}_{4} - \bar{\delta}\delta\bar{\Psi}_{3} \nonumber \\
    &- 2 \bar{\lambda} \Delta\bar{\Psi}_{1} - 2 \bar{\nu} \delta\bar{\Psi}_{1} + 2 \bar{\nu} \text{D}\bar{\Psi}_{2} + (3 \bar{\pi} +  \tau) \Delta\bar{\Psi}_{2} + (\bar{\gamma}- \gamma  + 3 \bar{\mu}) \delta\bar{\Psi}_{2} + 3 \bar{\lambda} \bar{\delta}\bar{\Psi}_{2} \nonumber \\
    &+ (\gamma - \bar{\gamma} - 3 \bar{\mu}) \text{D}\bar{\Psi}_{3} + (2 \bar{\rho} - \rho - 2 \bar{\epsilon}  ) \Delta\bar{\Psi}_{3} + (\alpha - 3 \bar{\beta} +  \bar{\tau}) \delta\bar{\Psi}_{3} - (2 \bar{\alpha} + 4 \bar{\pi} + \tau) \bar{\delta}\bar{\Psi}_{3} \nonumber \\
    &+ (3 \bar{\beta}- \alpha  - \bar{\tau}) \text{D}\bar{\Psi}_{4} - \bar{\kappa} \Delta\bar{\Psi}_{4} + (4 \bar{\epsilon} +  \rho - \bar{\rho}) \bar{\delta}\bar{\Psi}_{4} \nonumber \\
    &+2 \bar{\Psi}_{0} \bar{\lambda} \bar{\nu} +2 \bar{\Psi}_{1} [  \bar{\lambda} (\gamma - \bar{\gamma}  - 3  \bar{\mu} ) + \bar{\nu} (2 \bar{\alpha}  - 2  \bar{\pi} -  \tau ) - \Delta\bar{\lambda} ] \nonumber \\
    &+3\bar{\Psi}_{2} [\bar{\lambda} ( 3 \bar{\beta} -   \bar{\tau} - \alpha  ) + \bar{\pi} ( 3 \bar{\mu} -  \gamma  +  \bar{\gamma}  ) +  \bar{\nu}(  \rho - 2  \bar{\rho}) + \bar{\mu} \tau  +  \Delta\bar{\pi} +  \bar{\delta}\bar{\lambda} ] \nonumber \\
    &+2 \bar{\Psi}_{3} [ 2 \bar{\kappa} \bar{\nu} 
     +(\bar{\epsilon}-  \bar{\rho} ) ( \gamma -  \bar{\gamma}  - 3  \bar{\mu})
    -   \rho (\bar{\gamma} + 2 \bar{\mu})  
    + \tau (\bar{\tau} -  \bar{\beta}) 
    + (\bar{\alpha} + 2  \bar{\pi} ) (\alpha  - 3  \bar{\beta} + \bar{\tau}) 
     \nonumber \\
    & \hspace{14.0mm}   -  \Delta\bar{\epsilon} + \Delta\bar{\rho} -  \bar{\delta}\bar{\alpha} - 2 \bar{\delta}\bar{\pi} ] \nonumber \\
    &+ \bar{\Psi}_{4} [ 
       \bar{\kappa} (\gamma  - \bar{\gamma}  - 3 \bar{\mu} )
    + \rho ( 4 \bar{\beta}   -  \bar{\tau} )
     +  \bar{\rho} (\alpha  - 3 \bar{\beta}  +  \bar{\tau} )
    + 4\bar{\epsilon} (3 \bar{\beta} - \bar{\tau} - \alpha   )  - \bar{\sigma} \tau  \nonumber \\
    & \hspace{10.0mm} - \Delta\bar{\kappa} + 4 \bar{\delta}\bar{\epsilon} - \bar{\delta}\bar{\rho} ]\,, \label{BZ_proj_12}
\end{align}
%%%
\begin{align}
B^{Z}_{(2)(2)} = &\Delta\Delta\Psi_{0} -  \Delta\delta\Psi_{1} -  \delta\Delta\Psi_{1} + \delta\delta\Psi_{2} \nonumber \\
    &+(2 \mu -7 \gamma + \bar{\gamma} ) \Delta\Psi_{0} + \nu \delta\Psi_{0} -  \bar{\nu} \bar{\delta}\Psi_{0} \nonumber \\
    &+\bar{\nu} \text{D}\Psi_{1} + (7 \tau - \bar{\alpha} + 3 \beta ) \Delta\Psi_{1} + (5 \gamma -  \bar{\gamma} - 3 \mu) \delta\Psi_{1} + \bar{\lambda} \bar{\delta}\Psi_{1} \nonumber \\
    &- \bar{\lambda} \text{D}\Psi_{2} - 5 \sigma \Delta\Psi_{2} + (\bar{\alpha} -  \beta - 6 \tau) \delta\Psi_{2} + 4 \sigma \delta\Psi_{3} \nonumber \\
    &+ \Psi_{0} [
    \mu( \mu- 7 \gamma + \bar{\gamma} )
     +  \nu (\bar{\alpha} -  \beta - 3 \tau)
    + \bar{\nu} (4 \alpha  - \pi)
    +4 \gamma (3 \gamma -  \bar{\gamma}) -  \lambda \bar{\lambda}  \nonumber \\
    &\hspace{10.0mm}  - 4 \Delta\gamma + \Delta\mu + \delta\nu ] \nonumber \\
    &+2 \Psi_{1} [  2 \nu \sigma 
    - \bar{\nu} (\epsilon  + 2  \rho)
     +  \bar{\lambda} (\pi - \alpha) 
     +  (\bar{\gamma}-2 \gamma) (\beta  + 2  \tau )
   +  (\mu-\gamma) (5 \tau - \bar{\alpha}  + 2 \beta)  
    \nonumber \\
    &\hspace{14.0mm}   +  \Delta\beta + 2 \Delta\tau +  \delta\gamma -\delta\mu ] \nonumber \\
    &+3 \Psi_{2} [ \kappa \bar{\nu} + \bar{\lambda} \rho + \sigma (3 \gamma  -  \bar{\gamma}  - 3 \mu ) + \tau (3 \tau -  \bar{\alpha}  + \beta )  -  \Delta\sigma -  \delta\tau ] \nonumber \\
    &+2 \Psi_{3} [- \kappa \bar{\lambda} + \sigma (\bar{\alpha} + \beta - 5 \tau) + \delta\sigma ] +2 \Psi_{4} \sigma^2 \nonumber \\
    &+ \text{D}\text{D}\bar{\Psi}_{4} - \text{D}\delta\bar{\Psi}_{3} - \delta\text{D}\bar{\Psi}_{3} + \delta\delta\bar{\Psi}_{2} \nonumber \\
    &-4 \bar{\lambda} \delta\bar{\Psi}_{1} + 5 \bar{\lambda} \text{D}\bar{\Psi}_{2} +  \sigma \Delta\bar{\Psi}_{2} + ( \bar{\alpha} - \beta + 6 \bar{\pi}) \delta\bar{\Psi}_{2} \nonumber \\
    &+ (\beta-3 \bar{\alpha}  - 7 \bar{\pi}) \text{D}\bar{\Psi}_{3} - \kappa \Delta\bar{\Psi}_{3} + ( \epsilon - 5 \bar{\epsilon} + 3 \bar{\rho}) \delta\bar{\Psi}_{3} - \sigma \bar{\delta}\bar{\Psi}_{3} \nonumber \\
    &+ (7 \bar{\epsilon} -\epsilon - 2 \bar{\rho}) \text{D}\bar{\Psi}_{4} - \bar{\kappa} \delta\bar{\Psi}_{4} +  \kappa \bar{\delta}\bar{\Psi}_{4} \nonumber \\
    &+2 \bar{\Psi}_{0} \bar{\lambda}^2 +2 \bar{\Psi}_{1} [\bar{\lambda}( \bar{\alpha}  +  \beta  - 5  \bar{\pi}) - \bar{\nu} \sigma - \delta\bar{\lambda} ] \nonumber \\
    &+3 \bar{\Psi}_{2} [\kappa \bar{\nu}
    +\bar{\lambda}(3 \bar{\epsilon} -\epsilon    - 3  \bar{\rho})
    +  \bar{\mu} \sigma
    + \bar{\pi}(\bar{\alpha}  - \beta  + 3 \bar{\pi})  +  \text{D}\bar{\lambda} +  \delta\bar{\pi} ] \nonumber \\
    &+2 \bar{\Psi}_{3} [
     2 \bar{\kappa} \bar{\lambda}
     -\kappa( 2  \bar{\mu} +  \bar{\gamma})  
     + \sigma (\bar{\tau}-  \bar{\beta})
    +( \bar{\rho}-\bar{\epsilon}) (2 \bar{\alpha}  -  \beta + 5 \bar{\pi}  )
    +(\epsilon  -2\bar{\epsilon}) (2  \bar{\pi}+ \bar{\alpha}) 
    \nonumber \\
    & \hspace{14.0mm}  -  \text{D}\bar{\alpha} - 2 \text{D}\bar{\pi} -  \delta\bar{\epsilon} + \delta\bar{\rho} ] \nonumber \\
    &+\bar{\Psi}_{4} [\kappa( 4 \bar{\beta} -  \bar{\tau})  + \bar{\kappa}( \beta  - \bar{\alpha}    - 3  \bar{\pi})
    + (\bar{\rho}- 4\bar{\epsilon})( \epsilon - 3 \bar{\epsilon}   +  \bar{\rho})
      - \sigma \bar{\sigma} \nonumber \\
    & \hspace{10.0mm} + 4 \text{D}\bar{\epsilon} - \text{D}\bar{\rho} - \delta\bar{\kappa} ]\,, \label{BZ_proj_22}
\end{align}
%%%
\begin{align}
B^{Z}_{(1)(1)} = &\Delta\Delta\Psi_{2} -  \Delta\delta\Psi_{3} -  \delta\Delta\Psi_{3} + \delta\delta\Psi_{4} \nonumber \\
    &-4 \nu \Delta\Psi_{1} + (\gamma + \bar{\gamma} + 6 \mu) \Delta\Psi_{2} + 5 \nu \delta\Psi_{2} -  \bar{\nu} \bar{\delta}\Psi_{2} \nonumber \\
    &+\bar{\nu} \text{D}\Psi_{3} + (3 \tau-\bar{\alpha} - 5 \beta ) \Delta\Psi_{3} - (3 \gamma +  \bar{\gamma} + 7 \mu) \delta\Psi_{3} + \bar{\lambda} \bar{\delta}\Psi_{3} \nonumber \\
    &- \bar{\lambda} \text{D}\Psi_{4} -  \sigma \Delta\Psi_{4} + (\bar{\alpha} + 7 \beta - 2 \tau) \delta\Psi_{4} \nonumber \\
    &+ 2 \Psi_{0} \nu^2+ 2 \Psi_{1} [ \nu (\gamma  -  \bar{\gamma}  - 5 \mu)  + \lambda \bar{\nu} -  \Delta\nu ] \nonumber \\
    &+3 \Psi_{2} [ \mu (\gamma  + \bar{\gamma}  + 3 \mu) + \nu(\bar{\alpha}  + 3 \beta - 3  \tau) - \lambda \bar{\lambda} -  \bar{\nu} \pi  + \Delta\mu + \delta\nu ] \nonumber \\
    &+2 \Psi_{3} [\bar{\nu}( \epsilon -  \rho)
    + \bar{\lambda}( \alpha + 2  \pi) 
    +  \gamma (2 \tau 
    -\bar{\alpha} - 4 \beta )
     +  \bar{\gamma} (\tau - \beta)  
    +  \mu(5 \tau 
      - 2 \bar{\alpha}  - 9 \beta ) + 2 \nu \sigma \nonumber \\
    &\hspace{14.0mm} - \Delta\beta +  \Delta\tau - \delta\gamma - 2 \delta\mu ] \nonumber \\
    &+\Psi_{4} [ \kappa \bar{\nu}
     + \bar{\lambda} (\rho- 4 \epsilon)  
    -  \sigma(\gamma  +  \bar{\gamma} + 3 \mu )
    + 4 \beta(3 \beta 
    + \bar{\alpha} )  + \tau ( \tau-  \bar{\alpha} - 7 \beta ) \nonumber \\
    &\hspace{10.0mm} -  \Delta\sigma + 4 \delta\beta -  \delta\tau ]+ c.c.\,, \label{BZ_proj_11}
\end{align}
where $c.c.$ denotes the complex conjugation. Finally, the Bach tensor can be constructed as
\begin{align}
B_{(0)(0)} =& B^{Z}_{(0)(0)}+\Phi_{20}\Psi_{0}+\Phi_{02}\bar{\Psi}_{0}-2\Phi_{10}\Psi_{1}-2 \Phi_{01}\bar{\Psi}_{1} + \Phi_{00}(\Psi_{2}+ \bar{\Psi}_{2}) \,, \label{B_proj_00}\\
B_{(0)(1)} =& B^{Z}_{(0)(1)} + \Phi_{21}\Psi_{1} + \Phi_{12}\bar{\Psi}_{1} - 2 \Phi_{11}(\Psi_{2}+\bar{\Psi}_{2}) + \Phi_{01}\Psi_{3} + \Phi_{10}\bar{\Psi}_{3} \,, \label{B_proj_01}\\
B_{(0)(2)} =& B^{Z}_{(0)(2)} + \Phi_{21}\Psi_{0} - 2\Phi_{11}\Psi_{1}  + \Phi_{01}(\Psi_{2}- 2 \bar{\Psi}_{2}) + \Phi_{02}\bar{\Psi}_{1}  + \Phi_{00}\bar{\Psi}_{3} \,, \label{B_proj_02}\\
B_{(1)(2)} =& B^{Z}_{(1)(2)} + \Phi_{22}\Psi_{1}+\Phi_{12}( - 2\Psi_{2}+ \bar{\Psi}_{2} ) + \Phi_{02}\Psi_{3} - 2 \Phi_{11}\bar{\Psi}_{3}+ \Phi_{10}\bar{\Psi}_{4} \,, \label{B_proj_12}\\
B_{(2)(2)} =& B^{Z}_{(2)(2)} + \Phi_{22}\Psi_{0} - 2\Phi_{12}\Psi_{1} + \Phi_{02}(\Psi_{2} +\bar{\Psi}_{2}) - 2\Phi_{01}\bar{\Psi}_{3} + \Phi_{00}\bar{\Psi}_{4} \,, \label{B_proj_22}\\
B_{(1)(1)} =& B^{Z}_{(1)(1)} + \Phi_{22}(\Psi_{2}+\bar{\Psi}_{2}) - 2 \Phi_{12} \Psi_{3} - 2\Phi_{21}\bar{\Psi}_{3} + \Phi_{02}\Psi_{4} + \Phi_{20}\bar{\Psi}_{4} \,.\label{B_proj_11}
\end{align}
Since $\boldsymbol{m}$ is a complex vector, we have, e.g., ${\bar{B}_{(0)(2)} = B_{(0)(3)}}$, and since the Bach tensor is trace-less, it holds ${B_{(0)(1)} = B_{(2)(3)}}$, actually also ${B^{Z}_{(0)(1)} = B^{Z}_{(2)(3)}}$.

\section{Applications\label{Sec:Applications}}
To illustrate efficiency of the above general approach we analyze scenarios corresponding to special algebraic properties of the Ricci and Weyl tensors, respectively. This implies a specific behavior of privileged null geodesic congruence defining the Kundt and/or Robinson--Trautman classes in terms of its twist, shear, and expansion. In such important cases we discuss algebraic structure of the Bach tensor.

\subsection{Restrictions following from a special form of the Ricci tensor\label{SubSec:Prop}}

Let us use NP formalism to prove Propositions 1.2 and 1.1 in \cite{PravdaPravdovaPodolskySvarc:2017} in four dimensions. The original proof using higher-dimensional NP formalism was cleverly based on the analysis of dominant boost weights, however, here we can proceed fully explicitly. 

\subsubsection{Proof of Proposition 1.2}

First, let us prove Proposition 1.2, namely that a vacuum solution to quadratic gravity with traceless Ricci type III, i.e., the Ricci tensor of the form (using the frame ${\{\bll, \bn,\bm^{(i)}\}}$, see \cite{PravdaPravdovaPodolskySvarc:2017})
\be
R_{ab} = \Lambda g_{ab} + \psi'_i (\ell_a m^{(i)}_b
+ m^{(i)}_a \ell_b) + \omega' \ell_a\ell_b\,,
\quad \psi'_i\psi'_i\not=0,
\ee
and aligned Weyl tensor of Petrov type II, or more special, is \emph{necessarily Kundt}\footnote{By the definition, the Kundt family of geometries admits a non-twisting, shear-free, and non-expanding null geodesic congruence \cite{Kundt:1961, Kundt:1962}}.

Therefore, using the NP formalism notation, $\Psi_{0}=\Psi_{1}=0$ and the Ricci tensor is of the form (using the frame ${\{k^a,l^a,m^a,\bar m^a\}}$)
\be
R_{ab}=2\Phi_{22}k_a k_b-2\Phi_{12}(k_a \bar m_b+\bar m_a k_b)-2\Phi_{21}(k_a  m_b+ m_a k_b)+\Lambda g_{ab}\,,
\ee
where $\Lambda=$const., i.e., $\Phi_{00}=\Phi_{01} =\Phi_{10}=\Phi_{11}= \Phi_{02}=\Phi_{20}=0$.

Using the above assumption on the Petrov type II or III, the Bianchi identities imply $\kappa\Psi_{2}=0$ \eqref{Bianchi_a} or $\kappa\Psi_{3}=0$ \eqref{Bianchi_e}, respectively. For type III or N and $\Phi_{12}\not=0$, the Bianchi equations imply $\kappa\Phi_{12}=0$ \eqref{Bianchi_b}, while for $\Phi_{12}=0$ it follows that $\kappa\Phi_{22}=0$ \eqref{Bianchi_j}. Therefore, in all possible cases we obtain
\be
\kappa=0 \,,
\ee
and the multiple PND congruence generated by $\bk$ is necessary \emph{geodetic}.

Further, let us assume that the congruence is affinely parametrized and the frame is paralelly propagated along this congruence, i.e.,
\be
\epsilon=0\,,\qquad \pi=0\,.
\ee
Now, it is convenient to discuss specific Petrov types separately:
\begin{itemize}
\item Type II $(\Psi_{0}=\Psi_{1}=0)$: the QG field equation \eqref{QG_00} simplifies to $-4\mathfrak{a}B_{(0)(0)}^Z=0$, 
\begin{align}
B^{Z}_{(0)(0)}= &  \text{D}\text{D}\Psi_{2} 
-6 \rho \text{D}\Psi_{2} +3 \Psi_{2} ( 3 \rho^2 -  \sigma \bar{\sigma}  -  \text{D}\rho )  + c.c.=0\,, \label{BZ_proj_00_II}
\end{align}
and using the Ricci and Bianchi identities \eqref{Ricci_a} and \eqref{Bianchi_e} for type II it implies
\be
3\sigma\bar\sigma (\Psi_2+\bar\Psi_2)=0 \,,
\ee
and therefore
\be
\sigma=0\,.
\ee
Equivalently, it immediately follows from \eqref{Bianchi_b} that ${3\sigma\Psi_2=0}$ and thus ${\sigma=0}$.

The field equation (\ref{QG_02}) reduces to ${-4\mathfrak{a}B_{(0)(2)}^Z=0}$ which gives
\begin{align}
B^{Z}_{(0)(2)} =&  \text{D}\delta\Psi_{2} - 3 \tau \text{D}\Psi_{2}  - 3 \rho \delta\Psi_{2} -3 \Psi_{2} ( - 3 \rho \tau + \text{D}\tau)  - \delta\text{D}\bar{\Psi}_{2} - \text{D}\delta\bar{\Psi}_{2} + \text{D}\text{D}\bar{\Psi}_{3}\label{BZ_proj_02_II}\\
&+( \bar{\alpha} +  \beta ) \text{D}\bar{\Psi}_{2}  + 5 \bar{\rho} \delta\bar{\Psi}_{2}  - 4 \bar{\rho}\text{D}\bar{\Psi}_{3}  +3 \bar{\Psi}_{2} ( -  (\bar{\alpha} +  \beta )\bar{\rho}  + \delta\bar{\rho}) 
\nonumber \\ &
+2 \bar{\Psi}_{3} (  2 \bar{\rho}^2 - \text{D}\bar{\rho}) =0 \,.  \nonumber 
\end{align}
Using geometric identities \eqref{Ricci_a}, \eqref{Ricci_c}, \eqref{Ricci_k}, \eqref{Bianchi_e}, \eqref{Bianchi_g}, \eqref{Bianchi_h}, and \eqref{Bianchi_j} we obtain
\be
-4\rho\bar\rho\Phi_{12}=0 \,,
\ee
and therefore
\be
\rho=0 \,.
\ee
The spacetime has to belong necessarily to the \emph{Kundt} class.

\item Type III $(\Psi_{0}=\Psi_{1}=\Psi_{2}=0)$: the QG field equation \eqref{QG_00} is automatically satisfied, while equation \eqref{QG_02}, namely $-4\mathfrak{a}B_{(0)(2)}^Z=0$, reads
\be
B^{Z}_{(0)(2)} = 2 \sigma \text{D}\Psi_{3}
+2 \Psi_{3} ( - 3 \rho \sigma  + \text{D}\sigma) 
 + \text{D}\text{D}\bar{\Psi}_{3} 
 %\nonumber \\ &
- 4 \bar{\rho} \text{D}\bar{\Psi}_{3} 
+2 \bar{\Psi}_{3} (  2 \bar{\rho}^2 - \sigma \bar{\sigma} - \text{D}\bar{\rho} ) =0\,. \label{BZ_proj_02_III}
\ee
Using \eqref{Ricci_a}, \eqref{Ricci_b}, \eqref{Bianchi_g}, \eqref{Bianchi_h}, \eqref{Bianchi_j}, it implies (together with its complex conjugate)
\begin{align}
(\sigma\bar\sigma+\rho\bar\rho)\Phi_{12}+\sigma\rho\Phi_{21} &= 0\,, \\
(\sigma\bar\sigma+\rho\bar\rho)\Phi_{21}+\bar{\sigma}\bar{\rho}\Phi_{12} &= 0\,.
\end{align}
To have $\Phi_{12}\not=0$, the determinant should be vanishing, i.e.,
\be
(\sigma\bar\sigma)^2+(\rho\bar\rho)^2+\sigma\bar\sigma\rho\bar\rho=0 \,.
\ee
We thus get ${\sigma=0}$ and ${\rho=0}$ and the spacetime has to be \emph{Kundt}.

\item Type N $(\Psi_{0}=\Psi_{1}=\Psi_{2}=\Psi_{3}=0)$: the Bianchi identity \eqref{Bianchi_h} simplifies to ${-2\rho\Phi_{12}=0}$ and we immediately get that either ${\rho=0}$ or ${\Phi_{12}=0}$. Taking ${\rho=0}$, the Ricci identity \eqref{Ricci_a}, i.e., $D\rho=\rho^2+\sigma\bar\sigma$, implies ${\sigma=0}$. In the case $\Phi_{12}=0$ combined with the Weyl type N, the QG field equations \eqref{QG_00} and \eqref{QG_02} are automatically satisfied, while the equation \eqref{QG_22} becomes ${-4\mathfrak{a}B_{(2)(2)}^Z=0}$, namely
\be
B^{Z}_{(2)(2)} =  2 \Psi_{4} \sigma^2 + \text{D}\text{D}\bar{\Psi}_{4} 
- 2 \bar{\rho} \text{D}\bar{\Psi}_{4}+\bar{\Psi}_{4} (   \bar{\rho}^2 - \sigma \bar{\sigma}  - \text{D}\bar{\rho})=0\,. \label{BZ_proj_22_N}
\ee
Using \eqref{Ricci_a}, \eqref{Ricci_b}, \eqref{Bianchi_c}, \eqref{Bianchi_f}, \eqref{Bianchi_k} it takes the form
\be
-4\sigma^2\Psi_4=0
\ee
and we get ${\sigma=0}$. Moreover, a combination of the Bianchi identities \eqref{Bianchi_f} and \eqref{Bianchi_k} leads to ${\rho\Phi_{22}=\sigma\Psi_4}$ that gives
\be
\rho=0\,.
\ee
Therefore, in both cases the resulting spacetime has to be \emph{Kundt}.
\end{itemize}

\subsubsection{Proof of Proposition 1.1}

Now, let us explicitly prove Proposition 1.1 of \cite{PravdaPravdovaPodolskySvarc:2017}, namely that a vacuum solution to quadratic gravity with traceless Ricci type N, i.e., the Ricci tensor of the form
\be
R_{ab} = \Lambda g_{ab}  + \omega' \ell_a\ell_b\,, \quad \omega'\not=0\,,
\ee
and aligned Weyl tensor of Petrov type I, or more special, is necessarily Kundt.

Therefore, using the NP formalism, $\Psi_{0}=0$ and the Ricci tensor is of the form (using the frame ${\{k^a,l^a,m^a,\bar m^a\}}$)
\be
R_{ab}=2\Phi_{22}k_a k_b+\Lambda g_{ab}\,,
\ee
where $\Lambda=$const., i.e., $\Phi_{00}=\Phi_{01} =\Phi_{10}=\Phi_{11}= \Phi_{02}=\Phi_{20}=\Phi_{12}=\Phi_{21}=0$.

To prove this proposition let us begin with the Bianchi identity \eqref{Bianchi_j} which gives $\kappa\Phi_{22}=0$ and therefore
\be
\kappa=0 \,,
\ee
and the congruence is \emph{geodetic}. Further, let us assume that the congruence is affinely parametrized and the tetrad is paralelly propagated, i.e.,
\be
\epsilon=0\,, \qquad \pi=0\,.
\ee

\begin{itemize}
\item Type I: interestingly, in combination with geometric identities, the QG field equations \eqref{QG_00} and \eqref{QG_02} % $Z_{(0)(0)}=B_{(0)(0)}^Z=0$ $Z_{(0)(2)}=B_{(0)(2)}^Z=0$, 
are identically satisfied. The equation \eqref{QG_01} reduces to ${-4\mathfrak{a}B_{(0)(1)}^Z=0}$ and \eqref{BZ_proj_01} explicitly gives
\begin{align}
B^{Z}_{(0)(1)}= & \bar{\delta}\Delta\Psi_{1} - \text{D}\Delta\Psi_{2} -  \bar{\delta}\delta\Psi_{2} + \text{D}\delta\Psi_{3} \nonumber \\
&+2 \nu \text{D}\Psi_{1} + (- \alpha + \bar{\beta} ) \Delta\Psi_{1} + \lambda \delta\Psi_{1} + (-2 \gamma + 2 \mu -  \bar{\mu}) \bar{\delta}\Psi_{1} \nonumber \\
&+(-3 \mu + \bar{\mu}) \text{D}\Psi_{2} +  2 \rho \Delta\Psi_{2} + (\alpha -  \bar{\beta} ) \delta\Psi_{2} + 3 \tau \bar{\delta}\Psi_{2} \nonumber \\
&+(2 \beta  - 2 \tau) \text{D}\Psi_{3}  - 2 \rho \delta\Psi_{3} - 2 \sigma \bar{\delta}\Psi_{3} +\sigma \text{D}\Psi_{4} \nonumber \\
&+\Psi_{1} [2\gamma( \alpha  - \bar{\beta})  - 2 \lambda(\beta +2 \tau) +2 \mu(  \bar{\beta}-  \alpha ) + 2 \alpha \bar{\mu}  %\nonumber \\
%&\hspace{10.0mm}
  - 4 \nu \rho  + 2 \text{D}\nu - 2 \bar{\delta}\gamma + 2 \bar{\delta}\mu] \nonumber \\
&-3 \Psi_{2} [\rho(  \bar{\mu}- 2 \mu) -  \lambda \sigma + \tau(\alpha  -  \bar{\beta})   + \text{D}\mu -  \bar{\delta}\tau] \nonumber \\
&+2 \Psi_{3} [ 2 \rho( \tau - \beta )  + \sigma(\alpha  -  \bar{\beta})  + \text{D}\beta -  \text{D}\tau -  \bar{\delta}\sigma] \nonumber \\
&+\Psi_{4} ( - 2 \rho \sigma  + \text{D}\sigma) + c.c.\,, \label{BZ_proj_01_I}
\end{align}
which can be significantly simplified to
\be
(\rho\bar\rho+\sigma\bar\sigma)\Phi_{22}=0 \,.
\ee
This condition obviously implies
\be
\rho=0 \,, \qquad \sigma =0\,,
\ee
and the resulting spacetime has to be necessarily \emph{Kundt}.

\item Type II: employing ${\Psi_{0}=\Psi_{1}=0}$ the Bianchi identity \eqref{Bianchi_b} reduces to
\be
3\sigma\Psi_2=0
\ee
and therefore ${\sigma=0}$. Aternatively, the QG field equation \eqref{QG_00} becomes
\begin{align}
B^{Z}_{(0)(0)}= &  \text{D}\text{D}\Psi_{2} - 6 \rho \text{D}\Psi_{2}  +3 \Psi_{2} ( 3 \rho^2 -  \sigma \bar{\sigma}  -  \text{D}\rho )  + c.c.=0\,, \label{BZ_proj_00_IIN}
\end{align}
which gives
\be
3\sigma\bar\sigma(\Psi_2+\bar\Psi_2)=0 \,,
\ee
and therefore we get $\sigma=0$ again. The QG field equations \eqref{QG_02} and \eqref{QG_22} % $Z_{(0)(2)}=B_{(0)(2)}^Z=0$ $Z_{(2)(2)}=B_{(2)(2)}^Z=0$
are identically satisfied. However, the QG field equation \eqref{QG_01} implies%$Z_{(0)(1)}=B_{(0)(1)}^Z=0$
\be
-\rho\bar\rho (\Phi_{22}+\bar\Phi_{22})=0 \label{cond_II_01}
\ee
and therefore also  ${\rho=0}$. The spacetime has to be \emph{Kundt}. Since equation \eqref{cond_II_01} does not contain $\Psi_2$ it holds also for more algebraically special Petrov types III and N.

\item Type III: the Bianchi identity \eqref{Bianchi_h} implies
\be
2\sigma\Psi_3=0
\ee
and therefore ${\sigma=0}$. Employing equation \eqref{cond_II_01}, which does not contain $\Psi_2$ and it is valid also for Petrov type III, we and up with the \emph{Kundt} spacetime.

\item Type N: in this case, the last part of the Proposition 1.2 proof (discussing the subcase $\Phi_{12}=0$) can be used and therefore the spacetime is \emph{Kundt} again.
\end{itemize}

\subsection{The Bach tensor for Robinson--Trautman geometries of specific Weyl type \label{SubSec:RT_Bach_types}}

Let us examine possible Bach types for different Petrov types for Robinson--Trautman \cite{RobTra60, RobTra62} metrics
\be
\dd s=g_{uu} (u,r,x^i) \dd u^2 -2 \dd u\dd r+2 g_{ui} (u,r,x^i) \dd u\dd x^i + g_{ij} (u,r,x^i) \dd x^i \dd x^j\,, \label{RT}
\ee
admitting geodetic (${\kappa=0}$) shear-free (${\sigma=0}$), twist-free (${\rho=\bar\rho}$), and expanding (${\rho\neq 0}$) null congruence generated by
\be
\bk=\partial_r \,,
\ee
which is affinely parametrized ({$\epsilon+\bar\epsilon=0$}).  The coordinate $r$ is the affine parameter along the congruence, $u$ labels null hypersurfaces with $\bk$ tangent (normal), and ${x^2,\, x^3}$ cover the transverse Riemannian 2-space. Moreover without loss of generality, we employ a parallelly propagated frame, i.e.,
\begin{equation}
\pi= 0\,,\ \ {\epsilon=0}\,.
\end{equation}
In what follows, the Ricci equations \eqref{Ricci_a}--\eqref{Ricci_e}, \eqref{Ricci_k} will be useful, namely
\begin{align}
D\rho&= \rho^2+\Phi_{00}\,,\label{Drho}\\
D\tau&= \rho\tau+\Psi_1+\Phi_{01}\,,\label{Dtau}\\
D\alpha&=\rho\alpha+\Phi_{10}\,,\label{Dalpha}\\
D\beta&= \rho \beta+\Psi_1\,,\label{Dbeta} \\
D\gamma&= \alpha\tau+\beta\bar\tau+\Psi_2+\Phi_{11}-\frac{R}{24}\,, \\
\delta\rho&= \rho(\bar\alpha+\beta)-\Psi_1+\Phi_{01}\,.\label{deltarho}
\end{align}

\subsubsection{Petrov type N}

Let us start with the Petrov type N (with $\bk$ being PND) represented by the Weyl components
\be
\Psi_4\neq0\,, \qquad \Psi_0=\Psi_1=\Psi_2=\Psi_3=0\,.
\ee
Within this setting, the components of the Bach tensor components \eqref{B_proj_00}--\eqref{B_proj_11} simplify to
\begin{align}
B_{(0)(0)}=& 0 \ \ % (\mbox{b.w. +2}, \ % B_{00})
\\
B_{(0)(2)}=& \bar B_{(0)(2)}=B_{(0)(3)}= 0 \ \ \\ %?? (\mbox{b.w. +1}, \ B_{02}\,,\ B_{03})\\
B_{(0)(1)}=& B_{(2)(3)}=0 \ \ \\ %(\mbox{b.w. 0}, \ B_{01}\,,\ B_{22}\,,\ B_{33}\,,\ B_{23})\\
B_{(2)(2)}=& DD\bar\Psi_4-2\rho D\bar\Psi_4 \ \\ % (\mbox{b.w. 0})\\
B_{(1)(2)}=&-\bar B_{(1)(3)}=\bar\delta D\bar\Psi_4-(\alpha-3\bar\beta+\bar\tau) D\bar\Psi_4+\Phi_{10}\bar\Psi_4+\bar\Psi_4[\rho (\alpha+\bar\beta)-\bar\delta \rho]\\
%\nonumber\\
%=& -\bar\delta D\bar\Psi_4+2(\alpha-\bar\beta) D\bar\Psi_4-\Phi_{10}\bar\Psi_4+\bar\Psi_4[-\rho (\alpha+\bar\beta)+\bar\delta \rho]\ \ (\mbox{b.w. -1}, \ B_{12}\,,\ B_{13})\\
B_{(1)(1)}=& \delta\delta\Psi_4-\bar\lambda D\Psi_4+(\bar\alpha+7\beta-2\tau)\delta\Psi_4+\Phi_{02}\Psi_4\nonumber\\
& +\Psi_4[4\beta(\bar\alpha+3\beta)+\bar\lambda\rho+\tau(\tau-\bar\alpha-7\beta)+4\delta\beta-\delta\tau]+c.c.
%\nonumber\\
%=& -\delta\delta\Psi_4+\bar\lambda D\Psi_4-(-\bar\alpha+5\beta)\delta\Psi_4-\Phi_{02}\Psi_4
%%\nonumber\\ && 
%-\Psi_4[2\beta(3\beta-\bar\alpha)+\bar\lambda\rho+\delta(3\beta-\bar\alpha)]+c.c.\nonumber\\
%&
%\ \ (\mbox{b.w. -2}, \ B_{11})
\end{align}
Since the b.w. (boost weight) \emph{zero} component $B_{(2)(2)}$ is nonvanishing, the Bach tensor is in general of type~II. In a special case with ${B_{(2)(2)}=0}$, e.g., if $D\Psi_4=0$, then the Bach tensor becomes of type III. 

\subsubsection{Petrov type III}

For the Petrov type III, with the Weyl components
\be
\Psi_3\neq0\,,\ \Psi_4\neq0\,, \qquad \Psi_0=\Psi_1=\Psi_2=0\,,
\ee
the non-negative boost-weight components of the Bach tensor \eqref{B_proj_00}--\eqref{B_proj_11} simplify to
\begin{align}
	B_{(0)(0)}&= 0\,,\\
	B_{(0)(2)}&= DD\bar\Psi_3-4\rho D \bar\Psi_3-\bar\Psi_3 (\Phi_{00}-2\rho^2)\,,\\
	B_{(0)(1)}&= D\delta \Psi_3+2(\beta-\tau) D \Psi_3-2\rho\delta\Psi_3-\Psi_3[\Phi_{01}+2\rho(\beta-\tau)]+c.c.\,,\\
	B_{(2)(2)}&= DD\bar \Psi_4-D\delta\bar\Psi_3-\delta D\bar\Psi_3+(\beta-3\bar\alpha) D \bar\Psi_3+3\rho\delta\bar\Psi_3-2\rho D\bar\Psi_4-2\bar\Psi_3(\Phi_{01}-2\rho\bar\alpha)\,,
\end{align}
where we use the Ricci equations \eqref{Drho}--\eqref{Dbeta} and \eqref{deltarho}. Since the b.w. $+1$ component ${B_{(0)(2)}}$ is nonvanishing, the Bach tensor is of type I.

\subsubsection{Petrov type II/D}

For the Petrov type D, defined by
\be
\Psi_2\neq0\,, \qquad \Psi_0=\Psi_1=\Psi_3=\Psi_4=0\,,
\ee
even the highest b.w. $+2$ component, namely
\be
B_{(0)(0)}= DD\Psi_2-6\rho D \Psi_2-2\Psi_2 (\Phi_{00}-3\rho^2) +c.c.\,,
\ee
is nonvanishing and therefore, the Bach tensor is of general type G.

\section{Summary}

After a brief introduction of the quadratic gravity (\ref{QG_action}) and a suitable form of its field equations (\ref{QG_FEqs_separated_form}), we summarized basic definitions of the Newman--Penrose formalism, see section~\ref{Sec:Intro}. In the next section~\ref{Sec:QGconstraints}, we immediately proceed to our main result that is reformulation of the quadratic gravity field equations in terms of the NP quantities, see expressions (\ref{QG_00})--(\ref{QG_23}) with the substitution from (\ref{Z_proj_00})--(\ref{Z_proj_23}) and (\ref{BZ_proj_00})--(\ref{BZ_proj_11}), or appendix~\ref{App:FullFEs}. Interestingly, the Ricci tensor contribution to the field equations is only linear within this modified theory of gravity. Therefore, the procedure combining (\ref{QG_00})--(\ref{QG_23}) with the geometric constraints, listed in appendix~\ref{App:NP}, is thus similar as in the case of classic general relativity, i.e., we deal with the linear system of algebraic equations for the Ricci tensor frame components. The aim of these results is to provide a tool for analysis of (exact) solutions to the quadratic gravity, where the invariant assumptions on the algebraic properties of curavure tensors, or e.g., specific behavior of null geodesics, can be simply made. This should allow one to compare four-dimensional quadratic gravity with other theories of gravity, primarily with Einstein's general relativity, on the level of admitted solutions where the initial ansatz is introduced in terms of purely geometric conditions.

 In the subsequent section~\ref{Sec:Applications}, we present two simple examples of applicability of the above mentioned general expressions. In particular, its first subsection~\ref{SubSec:Prop} contains explicit calculations proving pair of propositions previously formulated in \cite{PravdaPravdovaPodolskySvarc:2017}, where the original proof was based on the highest boost-weights discussion which does not need knowledge of the complete Bach tensor. In the second subsection~\ref{SubSec:RT_Bach_types}, we analyze possible algebraic structure of the Bach tensor in the case of Robinson--Trautman geometries (\ref{RT}). The Weyl tensor is assumed to be of algebraically special Petrov type with respect to the frame associated with the privileged non-twisting, shear-free, and expanding null geodesic congruence. Under such conditions the admitted structure of the Bach tensor is discussed. These new results are summarized in the following table~\ref{tbl:Bach_types}.
\begin{table}[H]
	\begin{center}
		\begin{tabular}{|c|c|c|}
			\hline
			Petrov type
			& vanishing Bach components 
			%&possible non-vanishing Bach compts. 
			&
			possible Bach types
			\\[0.5mm]
			\hline\hline
			N & b.w. +2: $B_{(0)(0)}$, b.w. +1: $B_{(0)(2)}$, $B_{(0)(3)}$, b.w. 0: $B_{(0)(1)}$, $B_{(2)(3)}$ 
			%& $ B_{23}, B_{24}, B_{33}, B_{22}$ 
			&
			II/III/N/O
			\\[1mm]				
			\hline
			III & b.w. +2: $B_{(0)(0)}$
			% & $B_{12},B_{13},B_{14},B_{12}, B_{34}, B_{23}, B_{24}, B_{33}, B_{22}$ 
			&
			I/II/III/N/O
			\\[1mm]				
			\hline
			II/D & %& $B_{11},B_{12},B_{13}, B_{14},B_{12},  B_{34}, B_{23}, B_{24}, B_{33}, B_{22}$ 
			&
			G/I/II/III/III/N/O
			\\[1mm]				
			\hline
			\hline
		\end{tabular} \\[2mm]
		\caption{Possible Bach types depending on the Petrov type for Robinson--Trautman spacetimes. The privileged RT null vector field $\bk$ is taken as the Weyl PND.}
		\label{tbl:Bach_types}
	\end{center}
\end{table}

Moreover, the standard geometric Ricci and Bianchi identities of the Newman--Penrose formalism are summarized in appendix~\ref{App:NP} using unified notation of \cite{Stephanietal:2003}. For the readers convenience, subsequent appendix~\ref{App:Notation} compares this notation and conventions with other common textbooks \cite{Chandrasekhar:1993,PenroseRindler:1984}. Two decades ago the arbitrary-dimensional version of the Newman--Penrose formalism was introduced, and, almost immediately, it has become a useful tool with dozens of applications. Therefore, we present relation of such a real formalism, in the case of four spacetime dimensions, to the classic complex NP quantities used within this paper in appendix~\ref{App:Notation}. Finally, appendix~\ref{App:FullFEs} presents a fully explicit form of the quadratic gravity field equations, expressed in terms of the null frame quantities, which do not require any additional substitutions.

\section*{Acknowledgements}
AP is grateful for the support from the Czech Science Foundation Grant
No.~GA\v{C}R~19-09659S and the Research Plan RVO:
67985840. R\v{S} and DM were supported by the Czech Science Foundation grant No. GA\v{C}R 20-05421S and Charles University Grant Agency project No. 358921. 

\appendix

\section{Geometric constrains on the frame components}\label{App:NP}

In sections \ref{Sec:Intro} and \ref{Sec:QGconstraints} we have introduced frame components of crucial tensor quantities and constraints implied by the quadratic gravity field equations, respectively. In addition, these NP objects have to satisfy conditions directly arising from their purely geometric properties. In particular, we have commutation relations of the frame derivatives, the Ricci identities defining the Riemann tensor, and the Bianchi identities coming from the covariant derivatives of the Riemann tensor. For more details see \cite{Stephanietal:2003}.

\subsection{Commutation relations}

Expressing the Lie bracket of all possible combinations of the frame vectors, which are understood as the directional derivatives, and simultaneously, rewriting covariant derivatives in terms of the Ricci rotation coefficients we obtain the commutation relations, namely
\begin{align}
    \Delta \text{D}-\text{D} \Delta&=\left(\gamma+\bar{\gamma}\right) \text{D}+\left(\epsilon+\bar{\epsilon}\right) \Delta-\left(\bar{\tau}+\pi\right) \delta-\left(\tau+\bar{\pi}\right) \bar{\delta}\,, \label{commut_a} \\
    \delta \text{D}-\text{D} \delta &=\left(\bar{\alpha}+\beta-\bar{\pi}\right) \text{D}+\kappa \Delta-\left(\bar{\rho}+\epsilon-\bar{\epsilon}\right) \delta-\sigma \bar{\delta}\,, \label{commut_b}\\
    \delta \Delta-\Delta \delta &=-\bar{\nu} \text{D}+\left(\tau-\bar{\alpha}-\beta\right) \Delta+\left(\mu-\gamma+\bar{\gamma}\right) \delta+\bar{\lambda} \bar{\delta}\,, \label{commut_c} \\
    \bar{\delta} \delta-\delta \bar{\delta} &=\left(\bar{\mu}-\mu\right) \text{D}+\left(\bar{\rho}-\rho\right) \Delta+\left(\alpha-\bar{\beta}\right) \delta+\left(\beta-\bar{\alpha}\right) \bar{\delta}\,. \label{commut_d} 
\end{align}

\subsection{Ricci identities}

Using the notation of Ricci spin coefficients (\ref{NP_formalism_spin_coefficients}) the Riemann tensor nonzero components can be expressed as
\begingroup
\allowdisplaybreaks
\begin{align}
    \text{D}\sigma-\delta \kappa=&\,\sigma\left(3 \epsilon-\bar{\epsilon}+\rho+\bar{\rho}\right)+\kappa\left(\bar{\pi}-\tau-3 \beta-\bar{\alpha}\right)+\Psi_{0}\,,
    \label{Ricci_b} %\label{NP_field_equations_start}
     \\
    \text{D}\rho-\bar{\delta} \kappa=&\,\left(\rho^{2}+\sigma \bar{\sigma}\right)+\rho\left(\epsilon+\bar{\epsilon}\right) -\bar{\kappa} \tau + \kappa\left(\pi-3 \alpha-\bar{\beta}\right)+\Phi_{00}\,, \label{Ricci_a}\\
    \text{D}\tau-\Delta \kappa=&\,\rho\left(\tau+\bar{\pi}\right)+\sigma\left(\bar{\tau}+\pi\right)+\tau\left(\epsilon-\bar{\epsilon}\right) -\kappa\left(3 \gamma+\bar{\gamma}\right) +\Psi_{1}+\Phi_{01}\,,\label{Ricci_c} \\
    \text{D}\alpha-\bar{\delta} \epsilon=&\, \alpha\left(\rho+\bar{\epsilon}-2 \epsilon\right)+\beta \bar{\sigma}-\bar{\beta} \epsilon-\kappa \lambda -\bar{\kappa} \gamma+\pi(\epsilon+\rho)+\Phi_{10}\,,\label{Ricci_d} \\
    \text{D}\beta-\delta \epsilon=&\, \sigma(\alpha+\pi)+\beta\left(\bar{\rho}-\bar{\epsilon}\right)-\kappa(\mu+\gamma) +\epsilon\left(\bar{\pi}-\bar{\alpha}\right)+\Psi_{1}\,,\label{Ricci_e} \\
    \text{D}\gamma-\Delta \epsilon=&\, \alpha\left(\tau+\bar{\pi}\right)+\beta\left(\bar{\tau}+\pi\right)-\gamma\left(\epsilon+\bar{\epsilon}\right) -\epsilon\left(\gamma+\bar{\gamma}\right)+\tau \pi - \nu \kappa+\Psi_{2}+\Phi_{11}-\frac{1}{24}R\,, \label{Ricci_f} \\
    \text{D}\lambda-\bar{\delta} \pi=&\,\left(\rho \lambda+\bar{\sigma} \mu\right)+\pi(\pi+\alpha-\beta)-\nu \bar{\kappa} +\lambda\left(\bar{\epsilon}-3 \epsilon\right)+\Phi_{20}\,, \label{Ricci_g}\\
    \text{D}\mu-\delta \pi=&\,\left(\bar{\rho} \mu+\sigma \lambda\right)+\pi\left(\bar{\pi}-\bar{\alpha}+\beta\right) -\mu\left(\epsilon+\bar{\epsilon}\right)-\nu \kappa+\Psi_{2}+\frac{1}{12}R\,, \label{Ricci_h}\\
    \text{D}\nu-\Delta \pi=& \,\mu\left(\pi+\bar{\tau}\right)+\lambda\left(\bar{\pi}+\tau\right)+\pi\left(\gamma-\bar{\gamma}\right) -\nu\left(3 \epsilon+\bar{\epsilon}\right) +\Psi_{3}+\Phi_{21}\,,\label{Ricci_i} \\
    \Delta \lambda-\bar{\delta} \nu=&\, \lambda\left(\bar{\gamma}-3 \gamma-\mu-\bar{\mu}\right) +\nu\left(3 \alpha+\bar{\beta}+\pi-\bar{\tau}\right)-\Psi_{4}\,, \label{Ricci_j}\\
    \delta \rho-\bar{\delta} \sigma=&\, \rho\left(\bar{\alpha}+\beta\right)+\sigma\left(\bar{\beta}-3 \alpha\right) +\tau\left(\rho-\bar{\rho}\right) +\kappa\left(\mu-\bar{\mu}\right) -\Psi_{1}+\Phi_{01}\,, \label{Ricci_k}\\
    \delta \alpha-\bar{\delta} \beta=&\,(\mu \rho-\lambda \sigma)+\alpha \bar{\alpha}+\beta \bar{\beta}-2 \alpha \beta +\gamma\left(\rho-\bar{\rho}\right)+\epsilon\left(\mu-\bar{\mu}\right) -\Psi_{2}+\Phi_{11}+\frac{1}{24}R\,, \label{Ricci_l}\\
    \delta \lambda-\bar{\delta} \mu=&\, \nu \left(\rho-\bar{\rho}\right)+\pi\left(\mu-\bar{\mu}\right)+\mu\left(\alpha+\bar{\beta}\right) +\lambda\left(\bar{\alpha}-3 \beta\right) -\Psi_{3}+\Phi_{21}\,,\label{Ricci_m} \\
    \delta \nu-\Delta \mu=&\,\left(\mu^{2}+\lambda \bar{\lambda}\right)+\mu\left(\gamma+\bar{\gamma}\right)-\bar{\nu} \pi +\nu\left(\tau-3 \beta-\bar{\alpha}\right)+\Phi_{22}\,, \label{Ricci_n} \\
    \delta \gamma-\Delta \beta=&\, \gamma\left(\tau-\bar{\alpha}-\beta\right)+\mu \tau-\sigma \nu-\epsilon \bar{\nu} +\beta\left(\mu-\gamma+\bar{\gamma}\right)+\alpha \bar{\lambda} +\Phi_{12}\,, \label{Ricci_o} \\
    \delta \tau-\Delta \sigma=&\,\left(\mu \sigma+\bar{\lambda} \rho\right)+\tau\left(\tau+\beta-\bar{\alpha}\right) +\sigma\left(\bar{\gamma}-3 \gamma\right)-\kappa \bar{\nu}+\Phi_{02}\,, \label{Ricci_p}\\
    \Delta \rho-\bar{\delta} \tau=&\,-\left(\rho \bar{\mu}+\sigma \lambda\right)+\tau\left(\bar{\beta}-\alpha-\bar{\tau}\right) +\rho\left(\gamma+\bar{\gamma}\right)+\nu \kappa-\Psi_{2} -\frac{1}{12}R\,, \label{Ricci_q}\\
    \Delta \alpha-\bar{\delta} \gamma=&\, \nu(\rho+\epsilon)-\lambda(\tau+\beta)+\alpha\left(\bar{\gamma}-\bar{\mu}\right) +\gamma\left(\bar{\beta}-\bar{\tau}\right)-\Psi_{3}\,. \label{Ricci_r}
    %\label{NP_field_equations_end}
\end{align}
\endgroup

\subsection{Bianchi identities}
The projection of the Riemann tensor covariant derivative with cyclic exchange of indices leads to the first Bianchi identities,
\begingroup
\allowdisplaybreaks
\begin{align}
    0=&-\bar{\delta} \Psi_{0}+\text{D}\Psi_{1}+(4 \alpha-\pi) \Psi_{0}-2(2 \rho+\epsilon) \Psi_{1}+3 \kappa \Psi_{2} \nonumber \\
    &- \text{D}\Phi_{01}+\delta \Phi_{00}+2\left(\epsilon+\bar{\rho}\right) \Phi_{01}+2 \sigma \Phi_{10}-2 \kappa \Phi_{11}-\bar{\kappa} \Phi_{02} \nonumber \\
    &+\left(\bar{\pi}-2 \bar{\alpha}-2 \beta\right) \Phi_{00},\label{Bianchi_a} \\
    0=&+\bar{\delta} \Psi_{1}-\text{D}\Psi_{2}-\lambda \Psi_{0}+2(\pi-\alpha) \Psi_{1}+3 \rho \Psi_{2}-2 \kappa \Psi_{3} \nonumber \\
    &+ \bar{\delta} \Phi_{01}-\Delta \Phi_{00}-2\left(\alpha+\bar{\tau}\right) \Phi_{01}+2 \rho \Phi_{11}+\bar{\sigma} \Phi_{02} \nonumber \\
    &+\left(2 \gamma+2 \bar{\gamma} -\bar{\mu} \right) \Phi_{00}-2 \tau \Phi_{10}- \frac{1}{12}\text{D} R,\label{Bianchi_e} \\
    0=&-\bar{\delta} \Psi_{2}+\text{D}\Psi_{3}+2 \lambda \Psi_{1}-3 \pi \Psi_{2}+2(\epsilon-\rho) \Psi_{3}+\kappa \Psi_{4} \nonumber \\
    &- \text{D}\Phi_{21}+\delta \Phi_{20}+2\left(\bar{\rho}-\epsilon\right) \Phi_{21}-2 \mu \Phi_{10}+2 \pi \Phi_{11}-\bar{\kappa} \Phi_{22} \nonumber \\
    &+\left(2 \beta -2 \bar{\alpha} +\bar{\pi} \right) \Phi_{20}- \frac{1}{12}\bar{\delta} R,\label{Bianchi_g} \\
    0=&+\bar{\delta} \Psi_{3}-\text{D}\Psi_{4}-3 \lambda \Psi_{2}+2(2 \pi+\alpha) \Psi_{3}+ (\rho -4 \epsilon) \Psi_{4} \nonumber \\
    &- \Delta \Phi_{20}+\bar{\delta} \Phi_{21}+2\left(\alpha-\bar{\tau}\right) \Phi_{21}+2\nu \Phi_{10}+\bar{\sigma} \Phi_{22}-2 \lambda \Phi_{11} \nonumber \\
    &+\left(2 \bar{\gamma} -2 \gamma -\bar{\mu} \right) \Phi_{20},\label{Bianchi_c} \\
    0=&-\Delta \Psi_{0}+\delta \Psi_{1}+(4 \gamma-\mu) \Psi_{0}-2(2 \tau+\beta) \Psi_{1}+3 \sigma \Psi_{2} \nonumber \\
    &- \text{D}\Phi_{02}+\delta \Phi_{01}+2\left(\bar{\pi}-\beta\right) \Phi_{01}-2 \kappa \Phi_{12}-\bar{\lambda} \Phi_{00}+2 \sigma \Phi_{11} \nonumber \\
    &+\left(\bar{\rho}+2 \epsilon-2 \bar{\epsilon}\right) \Phi_{02},\label{Bianchi_b} \\
    0=&-\Delta \Psi_{1}+\delta \Psi_{2}+\nu\Psi_{0}+2(\gamma-\mu) \Psi_{1}-3 \tau \Psi_{2}+2 \sigma \Psi_{3} \nonumber \\
    &+\Delta \Phi_{01}-\bar{\delta} \Phi_{02}+2\left(\bar{\mu}-\gamma\right) \Phi_{01}-2 \rho \Phi_{12}-\bar{\nu} \Phi_{00}+2 \tau \Phi_{11} \nonumber \\
    &+\left(\bar{\tau}-2 \bar{\beta}+2 \alpha\right) \Phi_{02}+ \frac{1}{12}\delta R, \label{Bianchi_h}\\
    0=&-\Delta \Psi_{2}+\delta \Psi_{3}+2\nu \Psi_{1}-3 \mu \Psi_{2}+2(\beta-\tau) \Psi_{3}+\sigma \Psi_{4} \nonumber \\
    &- \text{D}\Phi_{22}+\delta \Phi_{21}+2\left(\bar{\pi}+\beta\right) \Phi_{21}-2 \mu \Phi_{11}-\bar{\lambda} \Phi_{20}+2 \pi \Phi_{12} \nonumber \\
    &+\left(\bar{\rho}-2 \epsilon-2 \bar{\epsilon}\right) \Phi_{22}- \frac{1}{12}\Delta R, \label{Bianchi_f}\\
    0=&-\Delta \Psi_{3}+\delta \Psi_{4}+3\nu \Psi_{2}-2(\gamma+2 \mu) \Psi_{3} +(4 \beta-\tau) \Psi_{4} \nonumber \\
    &+\Delta \Phi_{21}-\bar{\delta} \Phi_{22}+2\left(\bar{\mu}+\gamma\right) \Phi_{21}-2\nu \Phi_{11}-\bar{\nu} \Phi_{20}+2 \lambda \Phi_{12} \nonumber \\
    &+\left(\bar{\tau}-2 \alpha-2 \bar{\beta}\right) \Phi_{22} .\label{Bianchi_d}
    %\label{Bianchi_NP_first}
\end{align}
\endgroup
and the contraction gives the second Bianchi identities, namely
\begingroup
\allowdisplaybreaks
\begin{align}
    &\bar{\delta} \Phi_{01}+\delta \Phi_{10}-D\left(\Phi_{11}+ \frac{R}{8}\right)-\Delta \Phi_{00} \nonumber \\
    &\quad=\bar{\kappa} \Phi_{12}+\kappa \Phi_{21}+\left(2 \alpha+2 \bar{\tau}-\pi\right) \Phi_{01}+\left(2 \bar{\alpha}+2 \tau-\bar{\pi}\right) \Phi_{10} \nonumber \\
    &\quad-2\left(\rho+\bar{\rho}\right) \Phi_{11}-\bar{\sigma} \Phi_{02}-\sigma \Phi_{20}+\left[\mu+\bar{\mu}-2\left(\gamma+\bar{\gamma}\right)\right] \Phi_{00},\label{Bianchi_i} \\
    &\bar{\delta} \Phi_{12}+\delta \Phi_{21}-\Delta\left(\Phi_{11}+ \frac{R}{8}\right)-D \Phi_{22} \nonumber \\
    &=-\nu \Phi_{01}-\bar{\nu} \Phi_{10}+\left(\bar{\tau}-2 \bar{\beta}-2 \pi\right) \Phi_{12}+\left(\tau-2 \beta-2 \bar{\pi}\right) \Phi_{21} \nonumber \\
    &\quad+2\left(\mu+\bar{\mu}\right) \Phi_{11}+\left(2 \epsilon+ 2 \bar{\epsilon} -\rho-\bar{\rho} \right) \Phi_{22}+\lambda \Phi_{02}+\bar{\lambda} \Phi_{20}, \label{Bianchi_k}\\
    &\delta\left(\Phi_{11}- \frac{R}{8}\right)-D \Phi_{12}-\Delta\Phi_{01}+\bar{\delta} \Phi_{02} \nonumber \\
    &\quad=\kappa \Phi_{22}-\bar{\nu} \Phi_{00}+\left(\bar{\tau}-\pi+2 \alpha-2 \bar{\beta}\right) \Phi_{02}-\sigma \Phi_{21}+\bar{\lambda} \Phi_{10} \nonumber \\
    &\quad+2\left(\tau-\bar{\pi}\right) \Phi_{11}+\left( 2 \bar{\epsilon} -2 \rho-\bar{\rho}\right) \Phi_{12}+\left(2 \bar{\mu}+\mu-2 \gamma\right) \Phi_{01}.
    \label{Bianchi_j}
 %   \label{Bianchi_NP_second}
\end{align}
\endgroup

\section{Comparison of NP notation in classical textbooks}\label{App:Notation}

Within the geometric formulation of general relativity, several different conventions have appeared which typically affect signs of particular expressions. Here we follows the classic reference book \cite{Stephanietal:2003}, however, it is useful to compare our notation with other canonical sources \cite{Chandrasekhar:1993, PenroseRindler:1984}. The differences in notation\footnote{The definition of the energy-momentum tensor could be misleading in the Chandrasekhar book \cite{Chandrasekhar:1993}, namely, p. 34/ Eq. (236) gives ${G_{ij}=\frac{8\pi G}{c^4} T_{ij}}$ or, alternatively, 34/(236') is ${R_{ij}=\frac{8\pi G}{c^4} (T_{ij}-\pul T g_{ij})}$, however, p. 51/ Eq. (323) claims ${R_{ij}=-\frac{8\pi G}{c^4} (T_{ij}-\pul T g_{ij})}$ (for electromagnetic field see also p. 205/ Eq. (3) or p. 564/ Eq. (11)).} are summarized in table~\ref{definitions}.
\begin{table}[H]
	\begin{center}
		\begin{footnotesize}
		\begin{tabular}{|c|c|c|c|}
			\hline
		quantity
			& Stephani \cite{Stephanietal:2003}
			&	Chandrasekhar \cite{Chandrasekhar:1993}
			& Penrose, Rindler \cite{PenroseRindler:1984}
			\\[0.5mm]
			\hline\hline
		    signature & $+++-$ & $+---$ & $+---$
			\\[1mm]				
			\hline
		    frame & $m^a,\ \bar m^a,\ l^a,\ k^a$ &
		    $l^a,\ n^a,\ m^a,\ \bar m^a$ &
		     $l^a,\ n^a,\ m^a,\ \bar m^a$ 
			\\[1mm]				
			\hline
		Riemann t. & $ {R^a}_{bcd}=2{\Gamma^a}_{b[d,c]}+ 2{\Gamma^a}_{e[c}{\Gamma^e}_{d]b}$ &
		$ {R^a}_{bcd}=2{\Gamma^a}_{b[d,c]}+ 2{\Gamma^a}_{e[c}{\Gamma^e}_{d]b}$ &
		 $ {R^a}_{bcd}=-2{\Gamma^a}_{b[d,c]}- 2{\Gamma^a}_{e[c}{\Gamma^e}_{d]b}$
			\\[1mm]			
			\hline
			Einstein eqs. & $G_{ab}=\kappa T_{ab}$ & $G_{ab}=\pm\kappa T_{ab}$ &$G_{ab}=-\kappa T_{ab}$ 
			\\[1mm]			
			\hline
			NP compts. $\Psi_i$ & $\Psi_0=C_{abcd}k^am^b k^cm^d$\,, etc. &$\Psi_0=-C_{abcd}k^am^b k^cm^d$\,, etc. &$\Psi_0=C_{abcd}k^am^b k^cm^d$\,, etc.	\\[1mm]			
			\hline 
			NP compts. $\Phi_i$ &
			 $\Phi_{00}=\pul S_{ab}k^ak^b$\,, etc. & $\Phi_{00}=-\pul S_{ab}k^ak^b$\,, etc. & $\Phi_{00}=-\pul S_{ab}k^ak^b$\,, etc.	\\[1mm]			
			\hline
			Ricci rot. coeff. & $\kappa=-k_{a;b}m^a k^b$\,, etc.&$\kappa=k_{a;b}m^a k^b$\,, etc. & $\kappa=k_{a;b}m^a k^b$\,, etc.
			\\[1mm]			
			\hline
			\hline
		\end{tabular} \\[2mm]
		\caption{Notation comparison for the definitions of crucial geometric quantities.}
		\label{definitions}
		\end{footnotesize}
	\end{center}
\end{table}
However, the NP equations are the same in all three books \cite{Stephanietal:2003, Chandrasekhar:1993, PenroseRindler:1984}. To compare actual values of different quantities the subsequent table~\ref{values} can be used.
%\bea
%\ ^S g_{ab}&=& -\ ^C g_{ab}=-\ ^P g_{ab}\,,\\
%\ ^S k^a&=&+\ ^C \ell^a=+\ ^P \ell^a\,,\ 
%\ ^S l^a=+\ ^C n^a=+\ ^P n^a\,,
%\ ^S m^a=+\ ^C m^a=+\ ^P m^a\,,\ \\
%\ ^S k_a&=&-\ ^C \ell_a=-\ ^P \ell_a\,,\ 
%\ ^S l_a=-\ ^C n_a=-\ ^P n_a\,,
%\ ^S m_a=-\ ^C m_a=-\ ^P m_a\,,\ \\
%\ ^S {\Gamma^a}_{bcd}&=& +\ ^C {\Gamma^a}_{bcd}=+\ ^P {\Gamma^a}_{bcd}\,,\\
%\ ^S {R^a}_{bcd}&=& +\ ^C {R^a}_{bcd}
%=-\ ^P {R^a}_{bcd}\,,\\
%\eea

\begin{table}[H]
	\begin{center}
		\begin{footnotesize}
			\begin{tabular}{|c|c|}
				\hline
				quantity
				& values \\[0.5mm]
				\hline\hline
				metric & $\ ^S g_{ab}= -\ ^C g_{ab}=-\ ^P g_{ab}$
				\\[1mm]				
				\hline
				contravariant frame &$\ ^S k^a=+\ ^C \ell^a=+\ ^P \ell^a\,,\ 
				\ ^S l^a=+\ ^C n^a=+\ ^P n^a\,,
				\ ^S m^a=+\ ^C m^a=+\ ^P m^a$
				\\[1mm]				
				\hline
				covariant frame &
			$	\ ^S k_a=-\ ^C \ell_a=-\ ^P \ell_a\,,\ 
				\ ^S l_a=-\ ^C n_a=-\ ^P n_a\,,
				\ ^S m_a=-\ ^C m_a=-\ ^P m_a$
				\\[1mm]
				\hline
				Christoffel symbols &
				$\ ^S {\Gamma^a}_{bcd}= +\ ^C {\Gamma^a}_{bcd}=+\ ^P {\Gamma^a}_{bcd}$
				\\[1mm]
				\hline
				Riemann tensor & $ \ ^S {R^a}_{bcd}= +\ ^C {R^a}_{bcd}
				=-\ ^P {R^a}_{bcd}$ 
				\\[1mm]			
				\hline
					Weyl tensor & $ \ ^S {C^a}_{bcd}= +\ ^C {C^a}_{bcd}
				=-\ ^P {C^a}_{bcd}$ 
				\\[1mm]			
				\hline
				Ricci tensor &
					$\ ^S {R}_{ab}= +\ ^C {R}_{ab}=-\ ^P {R}_{ab}$
					\\[1mm]
					\hline
				Ricci scalar &
					$\ ^S {R}= -\ ^C {R}=+\ ^P {R}$
				\\[1mm]
				\hline
					Einstein tensor &
				$\ ^S {G}_{ab}= +\ ^C {G}_{ab}=-\ ^P {G}_{ab}$
				\\[1mm]
				\hline	
				cosm. const &
				$\ ^S {\Lambda}= -\ ^C {\Lambda}=+\ ^P {\Lambda}$
				\\[1mm]
				\hline
					stress-energy tensor &
				$\ ^S {T}_{ab}= \pm\ ^C {T}_{ab}=\ ^P {T}_{ab}$
				\\[1mm]
				\hline
				NP compts. $\Psi_i$ & $\ ^S\Psi_i=+\ ^C\Psi_i=+\ ^P\Psi_i$	\\[1mm]			
				\hline 				
				NP compts. $\Phi_i$ &
				$\ ^S\Phi_{ij} ={-}\ ^C \Phi_{ij}=+\ ^P \Phi_{ij}$ 	\\[1mm]			
				\hline
				Ricci rot. coefficients & $\ ^S\kappa=+ \ ^C\kappa=+ \ ^P\kappa $\,, etc.
				\\[1mm]			
				\hline
				\hline
			\end{tabular} \\[2mm]
			\caption{Values comparison summarized according to classic books by Stephani (S), Chandrasekhar (C), and Penrose, Rindler (P). {Note that the definition of $^C T_{ab}$ is not clear, see footnote 4. Also the sign in front of $^C \Phi_{ij}$ does not seem to be correct since the NP equations in all three books are same, i.e. the correct sigh should be $+$.}}
			\label{values}
		\end{footnotesize}
	\end{center}
\end{table}
From table~\ref{values} it follows that all scalars as defined in Chandrasekhar/Penrose books, appearing in the NP equations, have the opposite/same as in Stephani, respectively, and thus all NP equations have the same form.

\section{Comparison with HD NP formalism}
\label{App:ComparisonHD}

Since the computer implementation of symbolical calculation within classic Newman--Penrose formalism may become more difficult due to the presence of complex quantities, it can useful to employ its real version following from the real higher-dimensional (HD) NP formalism introduced in \cite{Coleyetal04, Durkeeetal10}, see also \cite{OrtPraPra13rev} for a 
review.\footnote{Note that in \cite{OrtPraPra07}, there are some sigh errors whenever there is an odd number of $\bn$'s in the expression, i.e., involving $\Delta$, $\phi$, $\psi'_i$, $\Psi_i$, $\Phi_{ij}$, $\Phi^A_{ij}$, $\Psi'_{ijk}$, $\tau_i$, $\tau'_i$, $\rho'_{ij}$,	$L_{10}$, $L_{1i}$, and ${M^{i}}_{j1}$.} Therefore, as a by-product, we derived relations between complex four-dimensional and real higher-dimensional NP formalisms. These identifications are presented in the form of tables~\ref{tab_frames}--\ref{tab_Ricci_rot_II_relation}. As abbreviation for the frame components of an arbitrary tensor $T_{\dots a\dots}$ let us use $T_{\dots (0)...}=T_{\dots a \dots} \ell^a$, $T_{\dots (1)\dots}=T_{\dots a\dots} n^a$, and $T_{\dots (i)\dots}=T_{\dots a\dots} m^a_{(i)}$. Moreover,  each index $T_{\dots (0)...}$, $T_{\dots (1)...}$, and $T_{\dots (i)...}$ contributes $+1$, $-1$, and $0$ to the boost weight of a component, respectively.

\begin{table}[H]
	\begin{center}
		%\begin{footnotesize}
			\begin{tabular}{|c|l|l|l|}
			\hline
				b.w. & $\quad +1$ & $\quad\quad\quad\quad\quad 0$ & $\quad -1$\\ \hline
 &	$\ell^a= k^a$ & $m^{(2)a}= \frac{1}{\sqrt{2}} (m^a+\bar m^a)$ & $n^a= -l^a$\\
	& & $m^{(3)a}= \frac{i}{\sqrt{2}} (m^a-\bar m^a)$ &  \\	\hline\hline
	&& $m^{a}= \frac{1}{\sqrt{2}} (m^{(2)a}-i m^{(3)a})$&\\		
  && $\bar m^{a}=\frac{1}{\sqrt{2}} (m^{(2)a}+i m^{(3)a})$&\\ \hline		
\end{tabular}
\caption{Relation between real (HD) frame vectors ${\{\bll,\, \bn,\,\bm^{(i)}\}}$ with $i=2,\,3$, satisfying $\ell_a n^a=1$, $m_{(i)}^a m^{(j)}_a=\delta^j_i$, and standard 4-dimensional NP frame vectors ${\{\bk,\, \bl,\, \bm,\, \bar\bm\}}$, see (\ref{null_condition_NP}).}
\label{tab_frames}
%\end{footnotesize}
\end{center}
\end{table}

%($D=\ell^a\nabla_a, \Delta=n^a\nabla_a,\delta^i=m^{a(i)}\nabla_a$) 
\begin{table}[H]
	\begin{center}
		%\begin{footnotesize}
		\begin{tabular}{|c|c|c|c|}
			\hline
			b.w. & $+1$ & $0$ & $-1$\\ \hline
			&	$D=\ell^a\nabla_a$ & $\delta^i=m^{a(i)}\nabla_a$ & $\Delta =n^a\nabla_a $\\
				\hline		
		\end{tabular}
		\caption{Definition of the directional derivatives in the HD NP notation.}
			\label{tab_derivatives_def}
		%\end{footnotesize}
	\end{center}
\end{table}

\begin{table}[H]
	\begin{center}
		%\begin{footnotesize}
			\begin{tabular}{|c|l|l|l|}
				\hline
				b.w. & $\quad +1$ & $\quad\quad\quad 0$ & $\quad -1$\\ \hline %\\[0.0001mm]
				&	$D=D$ & $\delta_2= \frac{\bar\delta+\delta }{\sqrt{2}}$ & $\Delta =-\Delta $\\
				& & $\delta_3 = \frac{i (\delta
					-\bar\delta )}{\sqrt{2}}$ &  \\	\hline\hline
				&&$\delta=  \frac{1}{\sqrt{2}} (\delta_2-i\delta_3)$&\\
				&&$\bar\delta=  \frac{1}{\sqrt{2}} (\delta_2+i\delta_3)$&\\ \hline		
			\end{tabular}
			\caption{Relation between directional derivatives in the classic NP formalism, i.e., $D=k^a\nabla_a$, $\Delta=l^a\nabla_a$, $\delta=m^{a}\nabla_a$, $\bar\delta=\bar m^{a}\nabla_a$, and in its HD reformulation.}
			\label{tab_derivatives_relation}
		%\end{footnotesize}
	\end{center}
\end{table}

\begin{table}[H]
	\begin{center}
		\begin{footnotesize}
			\begin{tabular}{|c|l|l|l|l|l|}
				\hline
				b.w.& $\quad \quad +2$  & $\quad\quad +1$ & $\quad\quad\ 0$ & $\quad\quad -1$ & $\quad\quad -2$\\ \hline
				& $\omega= R_{(0)(0)} $ & $\psi_i=R_{(0)(i)}$ & $\phi_{ij}=R_{(i)(i)}$ & $\psi'_{i}=R_{(1)(i)} $ &
				$\omega'=R_{(1)(1)}$  \\
				&& &$\phi=R_{(0)(1)}$&&\\
				\hline
			\end{tabular}
		\end{footnotesize}
		\caption{Definition of the HD Ricci components (frame ${\{\bll,\, \bn,\, \bm^{(i)}\}}$).}
			\label{tab_Ricci_def}
	\end{center}
\end{table}
%\begin{table}[H]
%	\begin{center}
%		\begin{footnotesize}
%			\begin{tabular}{|c|c|c|c|c|c|}
%				\hline
%				b.w.& $+2$  & $+1$ & $0$ & $-1$ & $-2$\\ \hline
%				& $\omega= R_{ab}\ell^a\ell^b $ & $\psi_i=R_{ab}\ell^a m^{(i)b}$ & $\phi_{ij}=R_{ab} m_{(i)}^a m_{(j)}^b$ & $\psi'_{i}=R_{ab} n^a m^{(i)b}$ &
%				$\omega'=R_{ab}n^a n^b$  \\
%				\hline
%			\end{tabular}
%		\end{footnotesize}
%	\end{center}
%\end{table}

\begin{table}[H]
	\begin{center}
		\begin{footnotesize}
		\begin{tabular}{|c|l|l|l|l|l|}
			\hline
			b.w.& $\quad +2$  & $\quad\quad\quad +1$ & $\quad\quad\quad\quad\quad\quad 0$ & $\quad\quad\quad\quad -1$ & $\quad -2$\\ \hline
			& $\omega= 2 \Phi_{00} $ & $\psi_2 = (\Phi_{01} + \bar\Phi_{01}) 
			\sqrt{2}$ & $\phi_{22} =\Phi_{02} + \bar\Phi_{02} + 
			2 \Phi_{11} + \frac{R}{4}$ & $\psi'_2 = -(\Phi_{12} + \bar\Phi_{12}) \sqrt{2}$ &
	    	$\omega'=  2 \Phi_{22}$ \\
			& & $\psi_3 = i (\Phi_{01} - \bar\Phi_{01}) \sqrt{2}$
			&$\phi_{33} =  -\Phi_{02} - \bar\Phi_{02} + 
			2 \Phi_{11} + \frac{R}{4}$
			 & $\psi'_3= -i (\Phi_{12} - \bar\Phi_{12}) \sqrt{2}$ & \\
			 & & & $\phi_{23}= i (\Phi_{02} - \bar\Phi_{02})$ & & \\
			  & & & $\phi =  -2 \Phi_{11} + \frac{R}{4}$ & & \\
			  \hline\hline
			  	& $ \Phi_{00}=\frac{\omega}{2}$ 
			  	& $\Phi_{01}= \frac{\psi_2-i\psi_3}{2\sqrt{2}}$ 
			  	& $\Phi_{02}=\frac{1}{4}(\phi_{22}-\phi_{33}-2i\phi_{23})$ 
			  	& $\Phi_{12}=-\frac{1}{2\sqrt{2}}(\psi'_2-i\psi'_3)$ 
			  	&
			  $ \Phi_{22}=\frac{\omega'}{2}$ \\
			  & & 
			  &$\Phi_{11}=\frac{1}{8}(-2\phi+\phi_{22}+\phi_{33})$
			  & & \\
			   & & & $R=2\phi+ \phi_{ii}$ & & \\
			  \hline
		  			\end{tabular}
		\end{footnotesize}
		\caption{Relation between Ricci components in the classic four-dimensional NP formalism and their HD counterparts, see also table~\ref{tab_Ricci_def}.}
		\label{tab_Ricci_relation}
	\end{center}
\end{table}

\begin{table}[H]
	\begin{center}
		%\begin{tiny} %
		\begin{footnotesize}
			\begin{tabular}{|c|l|l|l|l|l|}
				\hline
				b.w. & $\quad\quad +2$  & $\quad\quad\quad +1$ & $\quad\quad\quad\quad\quad 0$ & $\quad\quad\quad -1$ & $\quad\quad -2$\\ \hline
				&$\!\!\!\!
				\Omega_{ij}=C_{(0)(i)(0)(j)} $% \!\!\!\!$
				& $\!\!\!\!\Psi_i=C_{(0)(1)(0)(i)}$
				& $\!\!\!\!\Phi_{ij}=C_{(0)(i)(1)(j)}$
				& $\!\!\!\!\Psi'_i=C_{(1)(0)(1)(i)}$&
				$  \!\!\!\!
				 \Omega'_{ij}=C_{(1)(i)(1)(j)} %\!\!\!\!
				 $ \\
				%-----------------------
				&$\!\!\!\!\Omega_{ii} =0$ 
				& $ \!\!\!\!
				\Psi_{ijk}=C_{(0)(i)(j)(k)}$ %\!\!\!\!$
				& $ \!\!\!\!
				\Phi^A_{ij}=\frac{1}{2}C_{(0)(1)(i)(j)}$ % \!\!$
				&$ \!\!\!\!
				\Psi'_{ijk}=C_{(1)(i)(j)(k)} $ % \!\!\!\!$ 
				& 
				$\!\!\!\!\Omega'_{ii} =0
				$\\
				%-------------------------------------
				&$ $ & $\!\!\!\!\Psi_i=\Psi_{kik}$ & $\!\!\!\!\Phi=C_{(0)(1)(0)(1)}=\Phi_{ii}$ & $\!\!\!\!\Psi'_i=\Psi'_{kik}$
				& $	$\\ 
				& & &  $ \!\!\!\!
				\Phi^S_{ij}=-\frac{1}{2}C_{(i)(k)(j)(k)} %\!\!\!\!
				$& &\\ \hline
			\end{tabular}
			%\end{tiny} 
		\end{footnotesize}
		\caption{Definition of the HD Weyl components (frame ${\{\bll,\, \bn,\, \bm^{(i)}\}}$). In four dimensions, the Weyl tensor symmetries imply $\Omega_{33}=-\Omega_{22}$, $\Omega'_{33}=-\Omega'_{22}$, $\Phi^S_{22}=-\frac{1}{2}\Phi_{2323}=\Phi^S_{33}$, $\Phi^S_{23}=0=\Phi^S_{32}$, $\Psi_2=\Psi_{323}$, $\Psi_3=\Psi_{232}$, $\Psi'_2=\Psi'_{323}$, $\Psi'_3=\Psi'_{232}$.}
			\label{tab_Weyl_def}
	\end{center} 
\end{table}
%
%\begin{table}[H]
%	\begin{center}
%		\begin{tiny} %{footnotesize}
%			\begin{tabular}{|c|c|c|c|c|}
%				\hline
%				 $+2$  & $+1$ & $0$ & $-1$ & $-2$\\ \hline
%				 $\!\!\!\!\Omega_{ij}=C_{abcd}\ell^a m_{(i)}^b\ell^c m_{(j)}^d\!\!\!\!$
%				& $\Psi_i=C_{abcd}\ell^a n^b \ell^c m_{(i)}^d$
%				& $\Phi_{ij}=C_{abcd}\ell^a m_{(i)}^b n^c m_{(j)}^d$
%				& $\Psi'_i=C_{abcd}n^a \ell^b n^c m_{(i)}^d$&
%				$\!\!\!\!\Omega'_{ij}=C_{abcd}n^a m_{(i)}^b n^c m_{(j)}^d\!\!\!\!$ \\
%				%-----------------------
%				 $\Omega_{ii} =0$ 
%				& $\!\!\!\!\Psi_{ijk}=C_{abcd}\ell^a m_{(i)}^b m_{(j)}^c m_{(k)}^d\!\!\!\!$
%				& $\!\!\Phi^A_{ij}=\frac{1}{2}C_{abcd} \ell^a n^b m_{(i)}^c m_{(j)}^d\!\!$
%				&$\!\!\!\!\Psi'_{ijk}=C_{abcd}n^a m_{(i)}^b m_{(j)}^c m_{(k)}^d \!\!\!\!$ & 
%				$\Omega'_{ii} =0
%				$\\
%				%-------------------------------------
%				   $ $ & $\Psi_i=\Psi_{kik}$ & $\Phi=C_{abcd}\ell^a n^b \ell^c n^d=\Phi_{ii}$ & $\Psi'_i=\Psi'_{kik}$
%				& $	$\\ 
%				& &  $\!\!\!\!\Phi^S_{ij}=-\frac{1}{2}C_{abcd} m_{(i)}^a m_{(k)}^b m_{(j)}^c m_{(k)}^d\!\!\!\!$& &\\ \hline
%			\end{tabular}
%		\end{tiny} %{footnotesize}
%	\end{center} 
%\end{table}
 
\begin{table}[H]
	\begin{center}
		\begin{footnotesize}
			\begin{tabular}{|c|l|l|l|l|l|}
			\hline
			b.w.& $\quad \quad +2$  & $\quad\quad +1$ & $\quad\quad\quad 0$ & $\quad\quad -1$ & $\quad\quad -2$\\ \hline
			& $\!\!\!\!\Omega_{22}= \frac{1}{2}(\Psi_0+\bar\Psi_0)\!\!\!\!$
			  & $\!\!\!\!\Psi_2= -\frac{1}{\sqrt{2}}(\Psi_1+\bar\Psi_1)$
			  & $\!\!\!\!\Phi^S_{22}=  +\frac{1}{2}  (\Psi_2+\bar\Psi_2)\!\!\!\!$
			  & $\!\!\!\!\Psi'_2= \frac{1}{\sqrt{2}}(\Psi_3+\bar\Psi_3)\!\!\!\!$&
			   $\!\!\!\!\Omega'_{22}=\frac{1}{2}(\Psi_4+\bar\Psi_4)\!\!\!\!$ \\
			   %-----------------------
			 & $\!\!\!\!\Omega_{33} = -\frac{1}{2}(\Psi_0+\bar\Psi_0) %=-\Omega_{22}
			 \!\!\!\!$ 
			  & $\!\!\!\!\Psi_3 = -\frac{i}{\sqrt{2}} (\Psi_1-\bar\Psi_1)\!\!\!\!$
			  & $\!\!\!\!\Phi^S_{33}=\Phi^S_{22}$
			 &$\!\!\!\!\Psi'_3= -\frac{i}{\sqrt{2}} (\Psi_3-\bar\Psi_3)\!\!\!\!$ & 
			 $\!\!\!\!\Omega'_{33} = -\frac{1}{2} (\bar\Psi_4+\Psi_4)%=-\Omega'_{22}
			 \!\!\!\!$\\
			 %-------------------------------------
			&   $ \!\!\!\!\Omega_{23}=  \frac{i}{2}(\Psi_0-\bar\Psi_0) \!\!\!\!$ & & 
			$\!\!\!\!\Phi^S_{23}=0=\Phi^S_{32}$
			 &
			& $\!\!\!\!\Omega'_{23}=
			- \frac{i}{2} (\Psi_4-\bar\Psi_4)
			\!\!\!\!$\\
			&&&$\!\!\!\!\Phi^A_{23}=  -\frac{i}{2}  (\Psi_2-\bar\Psi_2)$&&\\
			&&&$\!\!\!\!\Phi=\bar\Psi_2+\Psi_2$&& \\ \hline
			\hline
			& $\!\!\!\!\Psi_0=\Omega_{22}-i\Omega_{23}$
			& $\!\!\!\!\Psi_1=-\frac{\Psi_2-i\Psi_3}{\sqrt{2}}$
			& $\!\!\!\!\Psi_2=\Phi^S_{22}+i\Phi^A_{23}$
			& $\!\!\!\!\Psi_3=\frac{\Psi'_2+i\Psi'_3}{\sqrt{2}}$
		   &$\!\!\!\!\Psi_4=\Omega'_{22}+i\Omega'_{23}$
			 \\
		\hline
		\end{tabular}
		\end{footnotesize}
		\caption{Relation between complex Weyl components in the four-dimensional NP formalism and their real HD counterparts, see also table~\ref{tab_Weyl_def}.}
		\label{tab_Weyl_relation}
	\end{center}
\end{table}
 
\begin{table}[H]
	\begin{center}
		\begin{footnotesize}
			\begin{tabular}{|c|l|l|l|l|l|}
				\hline
				b.w.& $\quad\quad +2$  & $\quad\quad +1$ & $\quad\quad 0$ & $\quad\quad -1$ & $\quad\quad -2$\\ \hline
				& $\kappa_i=\ell_{(i);(0)} $
				&$\rho_{ij}=\ell_{(i);(j)}$
				& $\tau_i=\ell_{(i);(1)} $
				& $ \rho'_{ij}=n_{(i);(j)} $
				&$  \kappa'_i=n_{(i);(1)} $
				\\
				& 
				& 
				& $\tau'_i=n_{(i);(0)}  $
				&
				& \\
				\hline
			\end{tabular}
		\end{footnotesize}
		\caption{Definition of the Ricci rotation coefficients in HD NP formalism with a specific boost weight.}
			\label{tab_Ricci_rot_I_def}
	\end{center}
\end{table}
%
%\begin{table}[H]
%	\begin{center}
%		\begin{footnotesize}
%			\begin{tabular}{|c|l|l|l|l|l|}
%				\hline
%				b.w.& $\ \ \ \ \  +2$  & $\ \ \ \ \ \ \ \ +1$ & $\ \ \  \ \ 0$ & $\ \ \ \ \ \ \ \ -1$ & $\ \ \ \ -2$\\ \hline
%				& $\kappa_i=\ell_{a;b}  m^a_{(i)} \ell^b$
%				&$\rho_{ij}=\ell_{a;b}  m^a_{(i)} m^b_{(j)}$
%				& $\tau_i=\ell_{a;b}  m^a_{(i)} n^b$
%				& $ \rho'_{ij}=n_{a;b}  m^a_{(i)} m^b_{(j)}$
%				&$  \kappa'_i=n_{a;b}  m^a_{(i)} n^b$
%				\\
%				& 
%				& 
%				& $\tau'_i=n_{a;b}  m^a_{(i)} \ell^b$
%				&
%				& \\
%				 \hline
%			\end{tabular}
%		\end{footnotesize}
%	\end{center}
%\end{table}

\begin{table}[H]
	\begin{center}
		\begin{footnotesize}
			\begin{tabular}{|c|l|l|l|}
				\hline
				b.w.&  $\quad\quad\quad +1$ & $\quad\quad \quad 0$ & $\quad\quad\quad -1$ \\ \hline
				& $L_{10}=\ell_{(1);(0)}  $
				& $L_{1i}=\ell_{(1);(i)}$
				& $L_{11}=\ell_{(1);(1)}  $
				\\
				& ${M^i}_{j0}=m^{(i)}_{(j);(0)}$
				&${M^i}_{jk}=m^{(i)}_{(j);(k)}  $
				& ${M^i}_{j1}=m^{(i)}_{(j);(1)}$
				\\
				\hline
			\end{tabular}
		\end{footnotesize}
		\caption{Definition of the Ricci rotation coefficients in HD NP formalism that have a boost weight only under constant boosts.}
			\label{tab_Ricci_rot_II_def}
	\end{center}
\end{table}
%\begin{table}[H]
%	\begin{center}
%		\begin{footnotesize}
%			\begin{tabular}{|c|l|l|l|}
%				\hline
%				b.w.&  $\ \ \ \ \ \ \ \ +1$ & $\ \ \  \ \ 0$ & $\ \ \ \ \ \ \ \ -1$ \\ \hline
%				& $L_{10}=\ell_{a;b}   n^a \ell^b$
%				& $L_{1i}=\ell_{a;b}   n^a m^b_{(i)}$
%				& $L_{11}=\ell_{a;b}   n^a n^b$
%				\\
%				& ${M^i}_{j0}=m^{(i)}_{a;b} m^a_{(j)} \ell^b$
%				&${M^i}_{jk}=m^{(i)}_{a;b} m^a_{(j)} m^b_{(k)} $
%				& ${M^i}_{j1}=m^{(i)}_{a;b} m^a_{(j)} n^b$
%				\\
%				\hline
%			\end{tabular}
%		\end{footnotesize}
%	\end{center}
%\end{table}

\begin{table}[H]
	\begin{center}
		\begin{scriptsize} %{footnotesize}
			\begin{tabular}{|c|l|l|l|l|l|}
			\hline
			b.w.& $\quad \quad  +2$  & $\quad\quad\quad\quad\quad +1$ & $\quad\quad\quad 0$ & $\quad\quad\quad\quad\quad -1$ & $\quad\quad -2$\\ \hline
			& $\!\!\!\!\kappa_2=  -\frac{1 }{\sqrt{2}}(\kappa +\bar\kappa)\!\!\!\!$
			&$\!\!\!\!\rho_{22}=  -\frac{1}{2} (\rho
			+\sigma +\bar\rho +\bar\sigma )\!\!\!\!$
			& $\!\!\!\!\tau_2=
			\frac{1 }{\sqrt{2}}(\tau +\bar\tau)\!\!\!\!$
			& $\!\!\!\!\rho'_{22} =- \frac{1}{2} (\lambda +\mu +\bar\lambda +\bar\mu )\!\!\!\!$
			&$\!\!\!\! \kappa'_2= \frac{1 }{\sqrt{2}}(\nu +\bar\nu)\!\!\!\!$
			\\
			& $\!\!\!\!\kappa_3= -\frac{i}{\sqrt{2}}(\kappa -\bar\kappa )\!\!\!\!$
			& $\!\!\!\!\rho_{33}=  \frac{1}{2} (\sigma-\rho+\bar\sigma -\bar\rho  )\!\!\!\!$
			& $\!\!\!\!\tau_3=  \frac{i }{\sqrt{2}}(\tau -\bar\tau )\!\!\!\!$
			& $\!\!\!\!\rho'_{33}= \frac{1}{2} (\lambda -\mu +\bar\lambda -\bar\mu )\!\!\!\!$
			& $\!\!\!\!\kappa'_3= - \frac{i }{\sqrt{2}}( \nu-\bar\nu )\!\!\!\!$\\
			& & 
			$\!\!\!\!\rho_{23}=  \frac{i}{2}  (\rho -\sigma
			-\bar\rho +\bar\sigma )\!\!\!\!$
			& $\!\!\!\!\tau'_2= -\frac{1}{\sqrt{2}}(\pi +\bar\pi )\!\!\!\!$
			& $\!\!\!\!	\rho'_{23}=  \frac{i}{2}  (\lambda -\mu-\bar\lambda +\bar\mu )\!\!\!\!$
			&\\
			&
			& $\!\!\!\!\rho_{32}=
		-	\frac{i}{2} (\rho +\sigma-\bar\rho -\bar\sigma  )\!\!\!\!$
			& $\!\!\!\!\tau'_3= \frac{i }{\sqrt{2}}(\pi-\bar\pi )\!\!\!\!$
			& $\!\!\!\!\rho'_{32}=  \frac{i}{2}  (\lambda +\mu -\bar\lambda
			-\bar\mu )\!\!\!\!$
			&\\ \hline
			\hline
			& $\!\!\!\!\kappa=-\frac{1 }{\sqrt{2}}(\kappa_2 -i\kappa_3)\!\!\!\!$
			&$\!\!\!\!\rho=-\frac{1}{2}(\rho_{22}+\rho_{33}+i(\rho_{23}-\rho_{32}))\!\!\!\!$
			& $\!\!\!\!\tau=\frac{1}{\sqrt{2}}(\tau_2-i\tau_3 )\!\!\!\!$
			&$\!\!\!\!\mu=-\frac{1}{2}(\rho'_{22}+\rho'_{33}+i(\rho'_{32}-\rho'_{23}))\!\!\!\!$
			&$\!\!\!\! \nu=\frac{1 }{\sqrt{2}}(\kappa'_2 +i\kappa'_3)\!\!\!\!$
			\\
			& 
			& $\!\!\!\!\sigma=-\frac{1}{2}(\rho_{22}-\rho_{33}-i(\rho_{23}+\rho_{32}))\!\!\!\!$
			& $\!\!\!\!\pi =-\frac{1 }{\sqrt{2}}(\tau'_2+i\tau'_3)\!\!\!\!$
			& $\!\!\!\!\lambda= -\frac{1}{2}(\rho'_{22}-\rho'_{33}+i(\rho'_{23}+\rho'_{32}))\!\!\!\!$
			& \\
			\hline
			\end{tabular}
		\end{scriptsize} %{footnotesize}
		\caption{Relation between Ricci rotation coefficients in the four-dimensional NP formalism and their real HD counterparts that transforms with a specific boost weight.}
		\label{tab_Ricci_rot_I_relation}
	\end{center}
\end{table}

\begin{table}[H]
	\begin{center}
		\begin{footnotesize}
			\begin{tabular}{|c|l|l|l|}
				\hline
				b.w.&  $\quad\quad\quad +1$ & $\quad\quad\quad\quad\quad\quad\quad\quad 0$ & $\quad\quad\quad -1$ \\ \hline
               & $L_{10}=  \varepsilon +\bar\varepsilon$
               & $L_{12}=\frac{1 }{\sqrt{2}}(\alpha +\beta+\bar\alpha +\bar\beta )$
               & $L_{11}= -(\gamma +\bar\gamma )$
              \\
               & ${M^2}_{30}= -i (\varepsilon -\bar\varepsilon )$
               & $L_{13}= \frac{i }{\sqrt{2}}(\beta-\alpha  +\bar\alpha -\bar\beta )$
               & ${M^2}_{31}=  i (\gamma-\bar\gamma )$
               \\
               & 
               & ${M^2}_{33}=  \frac{1}{\sqrt{2}}(\beta-\alpha -\bar\alpha +\bar\beta )$
               & 
                \\
               & 
               & ${M^2}_{32}= -\frac{i}{\sqrt{2}}(\alpha +\beta-\bar\alpha -\bar\beta 
               )$
               & \\
               \hline
               \hline
               & $\varepsilon=\frac{1}{2} (L_{10}+i {M^2}_{30})$
               & $\alpha=\frac{1}{2\sqrt{2}} (L_{12}+iL_{13}-{M^2}_{33}+i{M^2}_{32})$
               & $\gamma =-\frac{1}{2}(L_{11}+i{M^2}_{31})$
               \\
               & 
               & $\beta=\frac{1}{2\sqrt{2}} (L_{12}-iL_{13}+{M^2}_{33}+i{M^2}_{32})$
               & 
               \\
               \hline
\end{tabular}
\end{footnotesize}
\caption{Relation between Ricci rotation coefficients in the four-dimensional NP formalism and their real HD counterparts that transforms with a specific boost weight only under constant boosts.}
		\label{tab_Ricci_rot_II_relation}
\end{center}
\end{table}

%%%%%%%%%%%%%%%%%%%%%%%%%%

\section{Complete set of quadratic gravity field equations\label{App:FullFEs}}

Finally for the readers convenience and direct applicability, we list the fully explicit set of the quadratic gravity field equations (\ref{QG_FEqs}) expressed in terms of the null frame ${\{\bk\,, \bl\,, \bm\,, \bar{\bm}\}}$, see (\ref{null_condition_NP}). In fact, the below equations correspond to (\ref{QG_00})--(\ref{QG_23}) with $Z_{(a)(b)}$ substituted from (\ref{Z_proj_00})--(\ref{Z_proj_23}), where the quantities ${B^Z_{(a)(b)}}$ are substituted from (\ref{BZ_proj_00})--(\ref{BZ_proj_11}).

The $\bk\bk$-projection is
\begin{align}
0 =& -4\mathfrak{a} \left[\Phi_{20} \Psi_{0} + \Phi_{02} \bar{\Psi}_{0} - 2 \Phi_{10} \Psi_{1} - 2 \Phi_{01} \bar{\Psi}_{1} + \Phi_{00}( \Psi_{2} +  \bar{\Psi}_{2})\right] \nonumber \\
&+ 2\left(\frac{ 1}{\mathsf{k}} + 2\mathfrak{b}R \right)\Phi_{00} 
%\nonumber\\
%&
+ 2\mathfrak{b}\left[(\epsilon + \bar{\epsilon}) \text{D} R - \text{D}\text{D} R - \bar{\kappa} \delta R - \kappa \bar{\delta} R \right] 
\nonumber\\
&-4\mathfrak{a}
\Big[\bar{\delta}\bar{\delta}\Psi_{0} - \text{D}\bar{\delta}\Psi_{1} - \bar{\delta}\text{D}\Psi_{1} + \text{D}\text{D}\Psi_{2}
% \nonumber \\
%&
+\lambda \text{D}\Psi_{0} + \bar{\sigma} \Delta\Psi_{0} + ( 2 \pi-7 \alpha -  \bar{\beta} ) \bar{\delta}\Psi_{0} \nonumber \\
&+(5 \alpha + \bar{\beta} - 3 \pi) \text{D}\Psi_{1} -  \bar{\kappa} \Delta\Psi_{1} -  \bar{\sigma} \delta\Psi_{1} + (3 \epsilon + \bar{\epsilon} + 7 \rho) \bar{\delta}\Psi_{1} \nonumber \\
&-( \epsilon +  \bar{\epsilon} + 6 \rho) \text{D}\Psi_{2} + \bar{\kappa} \delta\Psi_{2} - 5 \kappa \bar{\delta}\Psi_{2} +4 \kappa \text{D}\Psi_{3} \nonumber \\
&+\Psi_{0} [\bar{\kappa} \nu+4\alpha( 3 \alpha +  \bar{\beta}) -  (\epsilon+  \bar{\epsilon} + 3  \rho) \lambda  + \pi(\pi- 7 \alpha  -  \bar{\beta} )  +\bar{\sigma}( \mu - 4 \gamma) 
%\nonumber \\
%&\hspace{10.0mm}  
+ \text{D}\lambda - 4 \bar{\delta}\alpha + \bar{\delta}\pi] \nonumber \\
&+2 \Psi_{1} [2 \kappa \lambda+ \bar{\kappa}(\gamma- \mu )+\rho (5\pi-9 \alpha  - 2 \bar{\beta})  + \bar{\sigma}(\beta  + 2 \tau) + \epsilon (2 \pi-4 \alpha - \bar{\beta}) +  \bar{\epsilon}(\pi-\alpha )   \nonumber \\
&\hspace{14.0mm}  +  \text{D}\alpha - \text{D}\pi +  \bar{\delta}\epsilon + 2 \bar{\delta}\rho] \nonumber \\
&+3 \Psi_{2} [\kappa(3 \alpha  + \bar{\beta}  - 3  \pi ) -  \bar{\kappa} \tau + \rho (\epsilon + \bar{\epsilon}  + 3 \rho) -  \sigma \bar{\sigma} -  \text{D}\rho -  \bar{\delta}\kappa] \nonumber \\
&+2 \Psi_{3} [ \kappa(\epsilon -  \bar{\epsilon}  - 5  \rho )+ \bar{\kappa} \sigma + \text{D}\kappa] 
+2 \Psi_{4} \kappa^2 + c.c. \Big]
\,, \label{Full_QG_00}
\end{align}
the $\bk\bl$-projection is
\begin{align}
0 =& -4\mathfrak{a} \left[\Phi_{21} \Psi_{1} + \Phi_{12} \bar{\Psi}_{1} - 2 \Phi_{11} (\Psi_{2} + \bar{\Psi}_{2}) + \Phi_{01} \Psi_{3} + \Phi_{10} \bar{\Psi}_{3}\right] \nonumber \\ 
&
+2  \left(\frac{ 1}{\mathsf{k}} + 2\mathfrak{b}R \right)\Phi_{11}
+\frac{ 1}{\mathsf{k}}\left(
\frac{R}{4}-\Lambda\right)
%\nonumber\\
%&
 + 2\mathfrak{b}\Big[\, 
 \Delta\text{D} R 
 - \delta\bar{\delta} R 
 - \bar{\delta}\delta R
 - (\gamma + \bar{\gamma} - \mu - \bar{\mu}) \text{D} R \nonumber \\ 
&\quad - (\rho  + \bar{\rho} )\Delta R  + (\alpha  - \bar{\beta}  + \bar{\tau} )\delta R %\nonumber \\
%&\quad
 + (\bar{\alpha}  - \beta   + \tau) \bar{\delta} R \Big]
\nonumber\\
&-4\mathfrak{a}\Big[\bar{\delta}\Delta\Psi_{1} - \text{D}\Delta\Psi_{2} -  \bar{\delta}\delta\Psi_{2} + \text{D}\delta\Psi_{3} - \lambda \Delta\Psi_{0} -  \nu \bar{\delta}\Psi_{0} \nonumber \\
&+2 \nu \text{D}\Psi_{1} + (2 \pi- \alpha + \bar{\beta} ) \Delta\Psi_{1} + \lambda \delta\Psi_{1} + (2 \mu -  \bar{\mu} -2 \gamma ) \bar{\delta}\Psi_{1} \nonumber \\
&+(\bar{\mu} -3 \mu ) \text{D}\Psi_{2} + (2 \rho- \epsilon -  \bar{\epsilon} ) \Delta\Psi_{2} + (\alpha -  \bar{\beta} - 2 \pi) \delta\Psi_{2} + (\bar{\pi} + 3 \tau) \bar{\delta}\Psi_{2} \nonumber \\
&+(2 \beta -  \bar{\pi} - 2 \tau) \text{D}\Psi_{3} -  \kappa \Delta\Psi_{3} + (\epsilon + \bar{\epsilon} - 2 \rho) \delta\Psi_{3} - 2 \sigma \bar{\delta}\Psi_{3} +\sigma \text{D}\Psi_{4} + \kappa \delta\Psi_{4} \nonumber \\
&+\Psi_{0} [\lambda( 4 \gamma  -   \mu +  \bar{\mu}) + \nu(\alpha  -  \bar{\beta}  - 2  \pi )-  \bar{\delta}\nu] \nonumber \\
&+2\Psi_{1} [\gamma( \alpha  -  \bar{\beta}  - 2  \pi) -  \lambda(\beta 
+  \bar{\pi}+2  \tau)+\mu(\bar{\beta} -  \alpha  +   2 \pi) +  \bar{\mu}(\alpha  -  \pi ) +\nu( \epsilon  +  \bar{\epsilon}- 2  \rho ) \nonumber \\
&\hspace{10.0mm}  +  \text{D}\nu -  \bar{\delta}\gamma +  \bar{\delta}\mu] \nonumber \\
&+3 \Psi_{2} [  \kappa \nu+\mu( 2  \rho -\epsilon  - \bar{\epsilon} ) - \bar{\mu} \rho +  \pi \bar{\pi}  +  \lambda \sigma + \tau(2 \pi-\alpha  +  \bar{\beta}   ) - \text{D}\mu +  \bar{\delta}\tau] \nonumber \\
&+2 \Psi_{3} [\kappa (\bar{\mu}- 2  \mu -  \gamma) +\epsilon(\beta -   \tau  -   \bar{\pi}) + \bar{\epsilon} (\beta -  \tau )
+ \rho( \bar{\pi}  - 2 \beta+ 2  \tau) +  \sigma(\alpha  -  \bar{\beta} - 2 \pi) \nonumber \\
&\hspace{14.0mm}    + \text{D}\beta -  \text{D}\tau -  \bar{\delta}\sigma] \nonumber \\
&+\Psi_{4} [\kappa(4 \beta  -  \bar{\pi} -   \tau) +  \sigma (\epsilon+ \bar{\epsilon}  - 2 \rho  )+ \text{D}\sigma ] + c.c.\Big]
\label{Full_QG_01} \,, 
%\
\end{align}
the $\bk\bm$-projection is
\begin{align}
0 =& -4\mathfrak{a} \left[\Phi_{21} \Psi_{0} - 2 \Phi_{11} \Psi_{1} + \Phi_{02} \bar{\Psi}_{1} + \Phi_{01}( \Psi_{2} - 2  \bar{\Psi}_{2}) + \Phi_{00} \bar{\Psi}_{3}\right] \nonumber \\
&+ 2\left(\frac{ 1}{\mathsf{k}} + 2\mathfrak{b}R \right)\Phi_{01}  
%\nonumber\\
%&
+ 2\mathfrak{b}\left[\bar{\pi} \text{D} R  -\text{D}\delta R  - \kappa \Delta R  + (\epsilon  - \bar{\epsilon} )\delta R \right]
\nonumber \\
&
 -4\mathfrak{a}\Big[\bar{\delta}\Delta\Psi_{0} - \text{D}\Delta\Psi_{1} -  \bar{\delta}\delta\Psi_{1} + \text{D}\delta\Psi_{2} \nonumber \\
 &+\nu \text{D}\Psi_{0} + (\pi-3 \alpha + \bar{\beta} ) \Delta\Psi_{0} + (\mu -  \bar{\mu} -4 \gamma ) \bar{\delta}\Psi_{0} \nonumber \\
 &+(2 \gamma - 2 \mu + \bar{\mu}) \text{D}\Psi_{1} + (\epsilon -  \bar{\epsilon} + 3 \rho) \Delta\Psi_{1} + (3 \alpha -  \bar{\beta} -  \pi) \delta\Psi_{1}  + (2 \beta + \bar{\pi} + 4 \tau) \bar{\delta}\Psi_{1} \nonumber \\
 &-( \bar{\pi} + 3 \tau) \text{D}\Psi_{2} - 2 \kappa \Delta\Psi_{2} - ( \epsilon - \bar{\epsilon} + 3 \rho) \delta\Psi_{2} - 3 \sigma \bar{\delta}\Psi_{2} +2 \sigma \text{D}\Psi_{3} + 2 \kappa \delta\Psi_{3} \nonumber \\
 &+\Psi_{0} [(4\gamma-\mu) (3 \alpha  -  \bar{\beta}-  \pi)   + \bar{\mu}(4 \alpha -  \pi) +\nu( \bar{\epsilon} -  \epsilon - 3 \rho   )
 -  \lambda \bar{\pi}   
 % \nonumber \\
 %&\hspace{10.0mm}
 + \text{D}\nu - 4 \bar{\delta}\gamma + \bar{\delta}\mu] \nonumber \\
 &+2 \Psi_{1} [ 2 \kappa \nu +(\mu- \gamma) (\epsilon -  \bar{\epsilon} + 3  \rho) -  \bar{\mu}(2 \rho + \epsilon)  + ( \beta+2\tau) (\pi   -3 \alpha  +   \bar{\beta}) 
 +  \bar{\pi}(\pi - \alpha) \nonumber \\ 
 &\hspace{14.0mm}
 %\nonumber \\ 
 %   &\hspace{14.0mm}
 +  \text{D}\gamma - \text{D}\mu +  \bar{\delta}\beta + 2 \bar{\delta}\tau] \nonumber \\
 &+3 \Psi_{2} [ \kappa(\bar{\mu} -2 \mu ) +  \bar{\pi} \rho + \sigma(3 \alpha  - \bar{\beta}  - \pi ) + \tau(\epsilon  - \bar{\epsilon}  + 3 \rho ) - \text{D}\tau - \bar{\delta}\sigma] \nonumber \\
 &+2 \Psi_{3} [\kappa(2 \beta  -   \bar{\pi}- 2  \tau) +\sigma(\bar{\epsilon} -  \epsilon   - 3 \rho)  + \text{D}\sigma] + 2 \Psi_{4} \kappa \sigma \nonumber \\
 &+ \delta\delta\bar{\Psi}_{1} - \delta\text{D}\bar{\Psi}_{2} - \text{D}\delta\bar{\Psi}_{2} + \text{D}\text{D}\bar{\Psi}_{3} \nonumber \\
 &- 2 \bar{\lambda} \delta\bar{\Psi}_{0} + 3 \bar{\lambda} \text{D}\bar{\Psi}_{1} +  \sigma \Delta\bar{\Psi}_{1} +(4 \bar{\pi} - 3 \bar{\alpha} - \beta ) \delta\bar{\Psi}_{1} \nonumber \\
 &+( \bar{\alpha} +  \beta - 5 \bar{\pi}) \text{D}\bar{\Psi}_{2} - \kappa \Delta\bar{\Psi}_{2} + ( \epsilon - \bar{\epsilon} + 5 \bar{\rho}) \delta\bar{\Psi}_{2} - \sigma \bar{\delta}\bar{\Psi}_{2} \nonumber \\
 &+( 3 \bar{\epsilon}-\epsilon  - 4 \bar{\rho}) \text{D}\bar{\Psi}_{3} - 3 \bar{\kappa} \delta\bar{\Psi}_{3} +  \kappa \bar{\delta}\bar{\Psi}_{3} + 2 \bar{\kappa} \text{D}\bar{\Psi}_{4} \nonumber \\
 &+\bar{\Psi}_{0} [\bar{\lambda}(5 \bar{\alpha}  +  \beta  - 3  \bar{\pi}) - \bar{\nu} \sigma - \delta\bar{\lambda}] \nonumber \\
 &+2 \bar{\Psi}_{1} [\kappa \bar{\nu}+ \bar{\alpha}(\bar{\alpha} +  \beta ) +\bar{\pi}( 2 \bar{\pi} - 3 \bar{\alpha}  -  \beta)   - \bar{\lambda}( 4  \bar{\rho}+  \epsilon ) + \sigma(\bar{\mu}  -  \bar{\gamma} )  + \text{D}\bar{\lambda} -  \delta\bar{\alpha} + \delta\bar{\pi} ] \nonumber \\
 &+3 \bar{\Psi}_{2} [2 \bar{\kappa} \bar{\lambda} -  \kappa \bar{\mu} +  \bar{\pi}(\epsilon -  \bar{\epsilon})  + \bar{\rho}( 4 \bar{\pi}-  \bar{\alpha}  -  \beta )  + \sigma \bar{\tau} -  \text{D}\bar{\pi} + \delta\bar{\rho}] \nonumber \\
 &+2 \bar{\Psi}_{3} (\kappa  (\bar{\beta}  - \bar{\tau} ) +  \bar{\kappa} (\beta   - 4 \bar{\pi} ) - \sigma \bar{\sigma}
 +  (\bar{\rho} - \bar{\epsilon}) (\epsilon  -  \bar{\epsilon}  + 2 \bar{\rho})  +  \text{D}\bar{\epsilon} - \text{D}\bar{\rho} - \delta\bar{\kappa}) \nonumber \\
 &+\bar{\Psi}_{4} [ \bar{\kappa}(  5 \bar{\epsilon} -\epsilon   - 3  \bar{\rho}) +  \kappa \bar{\sigma} +  \text{D}\bar{\kappa} ]\Big]
 \,, \label{Full_QG_02}
 \end{align}
the $\bl\bl$-projection is
\begin{align}
0 =& -4\mathfrak{a} \bigl(\Phi_{22} (\Psi_{2} + \bar{\Psi}_{2}) - 2 \Phi_{12} \Psi_{3} - 2 \Phi_{21} \bar{\Psi}_{3} + \Phi_{02} \Psi_{4} + \Phi_{20} \bar{\Psi}_{4}\bigr) \nonumber \\ 
&+ 2\left(\frac{ 1}{\mathsf{k}} + 2\mathfrak{b}R \right)\Phi_{22}
%\nonumber\\
%& 
+ 2\mathfrak{b}\left[ - \Delta\Delta R  -(\gamma + \bar{\gamma}) \Delta R  +  \nu \delta R  +  \bar{\nu} \bar{\delta} R  \right]
\nonumber \\
&
-4\mathfrak{a}\Big[\Delta\Delta\Psi_{2} -  \Delta\delta\Psi_{3} -  \delta\Delta\Psi_{3} + \delta\delta\Psi_{4} \nonumber \\
&-4 \nu \Delta\Psi_{1} + (\gamma + \bar{\gamma} + 6 \mu) \Delta\Psi_{2} + 5 \nu \delta\Psi_{2} -  \bar{\nu} \bar{\delta}\Psi_{2} \nonumber \\
&+\bar{\nu} \text{D}\Psi_{3} + (3 \tau-\bar{\alpha} - 5 \beta ) \Delta\Psi_{3} - (3 \gamma +  \bar{\gamma} + 7 \mu) \delta\Psi_{3} + \bar{\lambda} \bar{\delta}\Psi_{3} \nonumber \\
&- \bar{\lambda} \text{D}\Psi_{4} -  \sigma \Delta\Psi_{4} + (\bar{\alpha} + 7 \beta - 2 \tau) \delta\Psi_{4} \nonumber \\
&+ 2 \Psi_{0} \nu^2+ 2 \Psi_{1} [ \nu (\gamma  -  \bar{\gamma}  - 5 \mu)  + \lambda \bar{\nu} -  \Delta\nu ] \nonumber \\
&+3 \Psi_{2} [ \mu (\gamma  + \bar{\gamma}  + 3 \mu) + \nu(\bar{\alpha}  + 3 \beta - 3  \tau) - \lambda \bar{\lambda} -  \bar{\nu} \pi  + \Delta\mu + \delta\nu ] \nonumber \\
&+2 \Psi_{3} [\bar{\nu}( \epsilon -  \rho)
+ \bar{\lambda}( \alpha + 2  \pi) 
+  \gamma (2 \tau 
-\bar{\alpha} - 4 \beta )
+  \bar{\gamma} (\tau - \beta)  
+  \mu(5 \tau 
- 2 \bar{\alpha}  - 9 \beta ) + 2 \nu \sigma \nonumber \\
&\hspace{14.0mm} - \Delta\beta +  \Delta\tau - \delta\gamma - 2 \delta\mu ] \nonumber \\
&+\Psi_{4} [ \kappa \bar{\nu}
+ \bar{\lambda} (\rho- 4 \epsilon)  
-  \sigma(\gamma  +  \bar{\gamma} + 3 \mu )
+ 4 \beta(3 \beta 
+ \bar{\alpha} )  + \tau ( \tau-  \bar{\alpha} - 7 \beta ) \nonumber \\
&\hspace{10.0mm} -  \Delta\sigma + 4 \delta\beta -  \delta\tau ]+ c.c.\Big]
 \,, \label{Full_QG_11}
\end{align}
the $\bl\bm$-projection is
\begin{align}
 0 =& -4\mathfrak{a} \left[\Phi_{22} \Psi_{1}+\Phi_{12}( - 2  \Psi_{2} +  \bar{\Psi}_{2}) + \Phi_{02} \Psi_{3} - 2 \Phi_{11} \bar{\Psi}_{3} + \Phi_{10} \bar{\Psi}_{4}\right] \nonumber \\ 
&+ 2\left(\frac{ 1}{\mathsf{k}} + 2\mathfrak{b}R \right)\Phi_{12}
%\nonumber\\
%& 
+ 2\mathfrak{b}\left[\bar{\nu} \text{D} R  -  \tau \Delta R  -  \Delta\delta R  + (\gamma   -  \bar{\gamma} )\delta R \right] \nonumber \\
&
 -4\mathfrak{a}\Big[\Delta\Delta\Psi_{1} -  \Delta\delta\Psi_{2} -  \delta\Delta\Psi_{2} + \delta\delta\Psi_{3} \nonumber \\
 &- 2 \nu \Delta\Psi_{0} + ( 4 \mu-3 \gamma + \bar{\gamma} ) \Delta\Psi_{1} + 3 \nu \delta\Psi_{1} -  \bar{\nu} \bar{\delta}\Psi_{1} \nonumber \\
 &+ \bar{\nu} \text{D}\Psi_{2} + (5 \tau- \bar{\alpha} -  \beta ) \Delta\Psi_{2} + (\gamma -  \bar{\gamma} - 5 \mu) \delta\Psi_{2} + \bar{\lambda} \bar{\delta}\Psi_{2} \nonumber \\
 &- \bar{\lambda} \text{D}\Psi_{3} - 3 \sigma \Delta\Psi_{3} + (\bar{\alpha} + 3 \beta - 4 \tau) \delta\Psi_{3} + 2 \sigma \delta\Psi_{4} \nonumber \\
 &+ \Psi_{0} [\nu ( 5 \gamma  -  \bar{\gamma}  - 3 \mu ) + \lambda \bar{\nu} -  \Delta\nu ] \nonumber \\
 &+ 2 \Psi_{1} [\nu ( \bar{\alpha} - 4  \tau)  + \bar{\nu}(\alpha  -   \pi)
 -  \lambda \bar{\lambda} +(\gamma-\mu) (\gamma -   \bar{\gamma}  - 2  \mu) %\nonumber \\
 %&\hspace{14.0mm} 
 - \Delta\gamma + \Delta\mu + \delta\nu ] \nonumber \\
 &+3\Psi_{2} [ \mu( 4  \tau-  \bar{\alpha}  -  \beta )+  \bar{\lambda} \pi -  \bar{\nu} \rho + 2 \nu \sigma+  \tau ( \bar{\gamma}   -  \gamma )   + \Delta\tau - \delta\mu ] \nonumber \\
 &+ 2 \Psi_{3} [  \kappa \bar{\nu} -   \sigma (\bar{\gamma} + 4 \mu) +\tau ( 2\tau -  \bar{\alpha} - 3 \beta ) + \beta (\bar{\alpha}  + \beta)+ \bar{\lambda} (\rho  -  \epsilon)  -  \Delta\sigma + \delta\beta -  \delta\tau ] \nonumber \\
 &+\Psi_{4} [- \kappa \bar{\lambda} + \sigma (\bar{\alpha} + 5 \beta - 3 \tau) + \delta\sigma ] \nonumber \\
 &- \Delta\text{D}\bar{\Psi}_{3} +  \Delta\delta\bar{\Psi}_{2} +  \bar{\delta}\text{D}\bar{\Psi}_{4} - \bar{\delta}\delta\bar{\Psi}_{3} \nonumber \\
 &- 2 \bar{\lambda} \Delta\bar{\Psi}_{1} - 2 \bar{\nu} \delta\bar{\Psi}_{1} + 2 \bar{\nu} \text{D}\bar{\Psi}_{2} + (3 \bar{\pi} +  \tau) \Delta\bar{\Psi}_{2} + (\bar{\gamma}- \gamma  + 3 \bar{\mu}) \delta\bar{\Psi}_{2} + 3 \bar{\lambda} \bar{\delta}\bar{\Psi}_{2} \nonumber \\
 &+ (\gamma - \bar{\gamma} - 3 \bar{\mu}) \text{D}\bar{\Psi}_{3} + (2 \bar{\rho} - \rho - 2 \bar{\epsilon}  ) \Delta\bar{\Psi}_{3} + (\alpha - 3 \bar{\beta} +  \bar{\tau}) \delta\bar{\Psi}_{3} - (2 \bar{\alpha} + 4 \bar{\pi} + \tau) \bar{\delta}\bar{\Psi}_{3} \nonumber \\
 &+ (3 \bar{\beta}- \alpha  - \bar{\tau}) \text{D}\bar{\Psi}_{4} - \bar{\kappa} \Delta\bar{\Psi}_{4} + (4 \bar{\epsilon} +  \rho - \bar{\rho}) \bar{\delta}\bar{\Psi}_{4} \nonumber \\
 &+2 \bar{\Psi}_{0} \bar{\lambda} \bar{\nu} +2 \bar{\Psi}_{1} [  \bar{\lambda} (\gamma - \bar{\gamma}  - 3  \bar{\mu} ) + \bar{\nu} (2 \bar{\alpha}  - 2  \bar{\pi} -  \tau ) - \Delta\bar{\lambda} ] \nonumber \\
 &+3\bar{\Psi}_{2} [\bar{\lambda} ( 3 \bar{\beta} -   \bar{\tau} - \alpha  ) + \bar{\pi} ( 3 \bar{\mu} -  \gamma  +  \bar{\gamma}  ) +  \bar{\nu}(  \rho - 2  \bar{\rho}) + \bar{\mu} \tau  +  \Delta\bar{\pi} +  \bar{\delta}\bar{\lambda} ] \nonumber \\
 &+2 \bar{\Psi}_{3} [ 2 \bar{\kappa} \bar{\nu} 
 +(\bar{\epsilon}-  \bar{\rho} ) ( \gamma -  \bar{\gamma}  - 3  \bar{\mu})
 -   \rho (\bar{\gamma} + 2 \bar{\mu})  
 + \tau (\bar{\tau} -  \bar{\beta}) 
 + (\bar{\alpha} + 2  \bar{\pi} ) (\alpha  - 3  \bar{\beta} + \bar{\tau}) 
 \nonumber \\
 & \hspace{14.0mm}   -  \Delta\bar{\epsilon} + \Delta\bar{\rho} -  \bar{\delta}\bar{\alpha} - 2 \bar{\delta}\bar{\pi} ] \nonumber \\
 &+ \bar{\Psi}_{4} [ 
 \bar{\kappa} (\gamma  - \bar{\gamma}  - 3 \bar{\mu} )
 + \rho ( 4 \bar{\beta}   -  \bar{\tau} )
 +  \bar{\rho} (\alpha  - 3 \bar{\beta}  +  \bar{\tau} )
 + 4\bar{\epsilon} (3 \bar{\beta} - \bar{\tau} - \alpha   )  - \bar{\sigma} \tau  \nonumber \\
 & \hspace{10.0mm} - \Delta\bar{\kappa} + 4 \bar{\delta}\bar{\epsilon} - \bar{\delta}\bar{\rho} ]\Big]
 \,,\label{Full_QG_12} 
 %\\
 \end{align}
the $\bm\bm$-projection is
\begin{align}
0 =& -4\mathfrak{a} \left[\Phi_{22} \Psi_{0} - 2 \Phi_{12} \Psi_{1} + \Phi_{02} (\Psi_{2} +  \bar{\Psi}_{2}) - 2 \Phi_{01} \bar{\Psi}_{3} + \Phi_{00} \bar{\Psi}_{4}\right] \nonumber \\ 
&+ 2\left(\frac{ 1}{\mathsf{k}} + 2\mathfrak{b}R \right)\Phi_{02} 
%\nonumber\\
%& 
+ 2\mathfrak{b}\left[\bar{\lambda} \text{D} R  -  \sigma \Delta R +( -  \bar{\alpha}   + \beta )\delta R  -  \delta\delta R \right]
\nonumber \\
&
-4\mathfrak{a}\Big[\Delta\Delta\Psi_{0} -  \Delta\delta\Psi_{1} -  \delta\Delta\Psi_{1} + \delta\delta\Psi_{2} \nonumber \\
&+(2 \mu -7 \gamma + \bar{\gamma} ) \Delta\Psi_{0} + \nu \delta\Psi_{0} -  \bar{\nu} \bar{\delta}\Psi_{0} \nonumber \\
&+\bar{\nu} \text{D}\Psi_{1} + (7 \tau - \bar{\alpha} + 3 \beta ) \Delta\Psi_{1} + (5 \gamma -  \bar{\gamma} - 3 \mu) \delta\Psi_{1} + \bar{\lambda} \bar{\delta}\Psi_{1} \nonumber \\
&- \bar{\lambda} \text{D}\Psi_{2} - 5 \sigma \Delta\Psi_{2} + (\bar{\alpha} -  \beta - 6 \tau) \delta\Psi_{2} + 4 \sigma \delta\Psi_{3} \nonumber \\
&+ \Psi_{0} [
\mu( \mu- 7 \gamma + \bar{\gamma} )
+  \nu (\bar{\alpha} -  \beta - 3 \tau)
+ \bar{\nu} (4 \alpha  - \pi)
+4 \gamma (3 \gamma -  \bar{\gamma}) -  \lambda \bar{\lambda}  \nonumber \\
&\hspace{10.0mm}  - 4 \Delta\gamma + \Delta\mu + \delta\nu ] \nonumber \\
&+2 \Psi_{1} [  2 \nu \sigma 
- \bar{\nu} (\epsilon  + 2  \rho)
+  \bar{\lambda} (\pi - \alpha) 
+  (\bar{\gamma}-2 \gamma) (\beta  + 2  \tau )
+  (\mu-\gamma) (5 \tau - \bar{\alpha}  + 2 \beta)  
\nonumber \\
&\hspace{14.0mm}   +  \Delta\beta + 2 \Delta\tau +  \delta\gamma -\delta\mu ] \nonumber \\
&+3 \Psi_{2} [ \kappa \bar{\nu} + \bar{\lambda} \rho + \sigma (3 \gamma  -  \bar{\gamma}  - 3 \mu ) + \tau (3 \tau -  \bar{\alpha}  + \beta )  -  \Delta\sigma -  \delta\tau ] \nonumber \\
&+2 \Psi_{3} [- \kappa \bar{\lambda} + \sigma (\bar{\alpha} + \beta - 5 \tau) + \delta\sigma ] +2 \Psi_{4} \sigma^2 \nonumber \\
&+ \text{D}\text{D}\bar{\Psi}_{4} - \text{D}\delta\bar{\Psi}_{3} - \delta\text{D}\bar{\Psi}_{3} + \delta\delta\bar{\Psi}_{2} \nonumber \\
&-4 \bar{\lambda} \delta\bar{\Psi}_{1} + 5 \bar{\lambda} \text{D}\bar{\Psi}_{2} +  \sigma \Delta\bar{\Psi}_{2} + ( \bar{\alpha} - \beta + 6 \bar{\pi}) \delta\bar{\Psi}_{2} \nonumber \\
&+ (\beta-3 \bar{\alpha}  - 7 \bar{\pi}) \text{D}\bar{\Psi}_{3} - \kappa \Delta\bar{\Psi}_{3} + ( \epsilon - 5 \bar{\epsilon} + 3 \bar{\rho}) \delta\bar{\Psi}_{3} - \sigma \bar{\delta}\bar{\Psi}_{3} \nonumber \\
&+ (7 \bar{\epsilon} -\epsilon - 2 \bar{\rho}) \text{D}\bar{\Psi}_{4} - \bar{\kappa} \delta\bar{\Psi}_{4} +  \kappa \bar{\delta}\bar{\Psi}_{4} \nonumber \\
&+2 \bar{\Psi}_{0} \bar{\lambda}^2 +2 \bar{\Psi}_{1} [\bar{\lambda}( \bar{\alpha}  +  \beta  - 5  \bar{\pi}) - \bar{\nu} \sigma - \delta\bar{\lambda} ] \nonumber \\
&+3 \bar{\Psi}_{2} [\kappa \bar{\nu}
+\bar{\lambda}(3 \bar{\epsilon} -\epsilon    - 3  \bar{\rho})
+  \bar{\mu} \sigma
+ \bar{\pi}(\bar{\alpha}  - \beta  + 3 \bar{\pi})  +  \text{D}\bar{\lambda} +  \delta\bar{\pi} ] \nonumber \\
&+2 \bar{\Psi}_{3} [
2 \bar{\kappa} \bar{\lambda}
-\kappa( 2  \bar{\mu} +  \bar{\gamma})  
+ \sigma (\bar{\tau}-  \bar{\beta})
+( \bar{\rho}-\bar{\epsilon}) (2 \bar{\alpha}  -  \beta + 5 \bar{\pi}  )
+(\epsilon  -2\bar{\epsilon}) (2  \bar{\pi}+ \bar{\alpha}) 
\nonumber \\
& \hspace{14.0mm}  -  \text{D}\bar{\alpha} - 2 \text{D}\bar{\pi} -  \delta\bar{\epsilon} + \delta\bar{\rho} ] \nonumber \\
&+\bar{\Psi}_{4} [\kappa( 4 \bar{\beta} -  \bar{\tau})  + \bar{\kappa}( \beta  - \bar{\alpha}    - 3  \bar{\pi})
+ (\bar{\rho}- 4\bar{\epsilon})( \epsilon - 3 \bar{\epsilon}   +  \bar{\rho})
- \sigma \bar{\sigma}% \nonumber \\ & \hspace{10.0mm}
 + 4 \text{D}\bar{\epsilon} - \text{D}\bar{\rho} - \delta\bar{\kappa} ]\Big]
 \,, \label{Full_QG_22}
\end{align}
the $\bm\bar{\bm}$-projection is
\begin{align}
0 =& -4\mathfrak{a} \left[\Phi_{21} \Psi_{1} + \Phi_{12} \bar{\Psi}_{1} - 2 \Phi_{11} (\Psi_{2} + \bar{\Psi}_{2}) + \Phi_{01} \Psi_{3} + \Phi_{10} \bar{\Psi}_{3}\right] \nonumber \\ 
&+ 2\left(\frac{ 1}{\mathsf{k}} + 2\mathfrak{b}R \right)\Phi_{11} + \frac{1}{\mathsf{k}}\left(\Lambda- \frac{R}{4}\right)
%\nonumber \\
%&
  + 2\mathfrak{b}\Big[\, (\gamma + \bar{\gamma} -  \bar{\mu}) \text{D} R %\nonumber \\ 
%&\quad 
- \text{D}\Delta R- \Delta\text{D} R
\nonumber \\ 
&\quad
 +( \rho-\epsilon   - \bar{\epsilon}   ) \Delta R  
  +(- \alpha   + \bar{\beta}+ \pi- \bar{\tau} )\delta R 
  %\nonumber \\
%&\quad
     + (\bar{\pi} - \tau )\bar{\delta} R  + \bar{\delta}\delta R \Big]
\nonumber \\
&
-4\mathfrak{a}\Big[
\bar{\delta}\Delta\Psi_{1} - \text{D}\Delta\Psi_{2} -  \bar{\delta}\delta\Psi_{2} + \text{D}\delta\Psi_{3} - \lambda \Delta\Psi_{0} -  \nu \bar{\delta}\Psi_{0} \nonumber \\
&+2 \nu \text{D}\Psi_{1} + (2 \pi- \alpha + \bar{\beta} ) \Delta\Psi_{1} + \lambda \delta\Psi_{1} + (2 \mu -  \bar{\mu} -2 \gamma ) \bar{\delta}\Psi_{1} \nonumber \\
&+(\bar{\mu} -3 \mu ) \text{D}\Psi_{2} + (2 \rho- \epsilon -  \bar{\epsilon} ) \Delta\Psi_{2} + (\alpha -  \bar{\beta} - 2 \pi) \delta\Psi_{2} + (\bar{\pi} + 3 \tau) \bar{\delta}\Psi_{2} \nonumber \\
&+(2 \beta -  \bar{\pi} - 2 \tau) \text{D}\Psi_{3} -  \kappa \Delta\Psi_{3} + (\epsilon + \bar{\epsilon} - 2 \rho) \delta\Psi_{3} - 2 \sigma \bar{\delta}\Psi_{3} +\sigma \text{D}\Psi_{4} + \kappa \delta\Psi_{4} \nonumber \\
&+\Psi_{0} [\lambda( 4 \gamma  -   \mu +  \bar{\mu}) + \nu(\alpha  -  \bar{\beta}  - 2  \pi )-  \bar{\delta}\nu] \nonumber \\
&+2\Psi_{1} [\gamma( \alpha  -  \bar{\beta}  - 2  \pi) -  \lambda(\beta 
+  \bar{\pi}+2  \tau)+\mu(\bar{\beta} -  \alpha  +   2 \pi) +  \bar{\mu}(\alpha  -  \pi ) +\nu( \epsilon  +  \bar{\epsilon}- 2  \rho ) \nonumber \\
&\hspace{10.0mm}  +  \text{D}\nu -  \bar{\delta}\gamma +  \bar{\delta}\mu] \nonumber \\
&+3 \Psi_{2} [  \kappa \nu+\mu( 2  \rho -\epsilon  - \bar{\epsilon} ) - \bar{\mu} \rho +  \pi \bar{\pi}  +  \lambda \sigma + \tau(2 \pi-\alpha  +  \bar{\beta}   ) - \text{D}\mu +  \bar{\delta}\tau] \nonumber \\
&+2 \Psi_{3} [\kappa (\bar{\mu}- 2  \mu -  \gamma) +\epsilon(\beta -   \tau  -   \bar{\pi}) + \bar{\epsilon} (\beta -  \tau )
+ \rho( \bar{\pi}  - 2 \beta+ 2  \tau) +  \sigma(\alpha  -  \bar{\beta} - 2 \pi) \nonumber \\
&\hspace{14.0mm}    + \text{D}\beta -  \text{D}\tau -  \bar{\delta}\sigma] \nonumber \\
&+\Psi_{4} [\kappa(4 \beta  -  \bar{\pi} -   \tau) +  \sigma (\epsilon+ \bar{\epsilon}  - 2 \rho  )+ \text{D}\sigma ] + c.c.
\Big]
 \,.
\label{Full_QG_23}
\end{align}

%%%%%%%%%%%%%%%%%%%%%%%%%%%%%%%%%%%%%%%%%%%%%%%%%%%%%%%%%%%%%%%%%%%%%%%%%%%%%%%%%

\end{document}